\documentclass[12pt]{article} 
\usepackage{tikz}
\usepackage{graphicx}
\tikzset{ lines/.style=very thin}
\usepackage{amssymb, amsmath, amsthm}
\usepackage[arrow,matrix,curve,cmtip,ps]{xy}
\usepackage{xcolor}
\newcommand{\red}{}
{
\begin{tabular}{|c|} \hline \\ $\displaystyle \ \ \ }%
{\ \ \ $\\ \\ \hline \end{tabular}} 

\newcommand{\tempout}[1]{{}}

\newcommand{\Ll}{{\mathcal{L}}}
\renewcommand{\L}{{\mathcal{L}}}
\renewcommand*{\dot}{\raisebox{-0.88ex}{\scalebox{2.5}{$\cdot$}}}
\newcommand{\A}{{\mathbf A}}

\newcommand{\E}{{\mathbf E}}
\newcommand{\Jor}{\E} 

\newcommand{\M}{\boldsymbol{\mathcal M}}

\newcommand{\V}{{\mathbf V}} 

\newcommand{\x}{\hat{x}}
\newcommand{\y}{\hat{y}}
\newcommand{\X}{\hat{X}}
\renewcommand{\u}{\hat{u}}
\renewcommand{\v}{\hat{v}}

\newcommand{\R}{{\mathbb{R}}}
\newcommand{\C}{{\mathbb{C}}}

\newcommand{\Q}{{\mathbb{H}}}
\newcommand{\F}{{\mathbb{F}}}
\newcommand{\Oct}{{\mathbb{O}}}

\renewcommand{\H}{\boldsymbol{\mathcal{H}}}

\newcommand{\Mat}{\mathbb{M}}
\newcommand{\sa}{\mbox{sa}}

\renewcommand{\1}{{\mathbf 1}}

\newcommand{\tr}{\mbox{Tr}}
\newcommand{\Tr}{\mbox{Tr}}

\renewcommand{\hat}{\widehat}
\newcommand{\id}{\mbox{id}}
\renewcommand{\bar}{\overline}

\newcommand{\Cat}{{\mathcal C}}

\newcommand{\EEJA}{\mathbf{EEJA}}

\newcommand{\FdHilbR}{\mathbf{FdHilb}_{\R}}

\newcommand{\op}{\text{op}}

\newcounter{thaler}
\newenvironment{mlist}{\begin{list}{\arabic{thaler}}%
{\usecounter{thaler}
\setlength{\rightmargin}{\leftmargin}
\topsep=0pt
\itemsep=0pt
\parskip=0pt
\parsep=0pt
}}{\end{list}}

\title{\bf A Royal Road to Quantum Theory \\(or Thereabouts)}
\author{Alexander Wilce\\Department of Mathematics, Susquehanna University} 
\date{\today} 
\begin{document}
\maketitle



\begin{abstract} 

This paper fails to derive quantum mechanics from a few simple postulates. But it gets very close, and does so without much exertion.  More precisely, I obtain a representation of finite-dimensional probabilistic systems in terms of euclidean Jordan algebras, in a strikingly easy way, from simple assumptions. 
This provides a framework within which real, complex and quaternionic QM can play happily together, and allows some --- but not too much --- room for more exotic alternatives. 
(This is a leisurely summary, based on recent lectures, of material from 
the papers arXiv:1206:2897 and arXiv:1507.06278, the latter joint work with Howard Barnum and Matthew Graydon. Some further ideas are also explored.)
\end{abstract}

\section{Introduction and Overview} 

Whatever else it may be, Quantum mechanics (QM) is a machine for making  {\em probabilistic} predictions about the results of measurements. To this extent, QM is, at least in part, about information. Recently, it has become clear that the formal apparatus of quantum theory, at least in finite dimensions, can be recovered from constraints on how physical systems store and process information. To this extent, finite-dimensional QM is {\em just} about information. 

The broad idea of regarding QM in this way, and of attempting to derive its mathematical structure from simple operational 
or probabilistic axioms, is not new. Efforts in this direction 
go back at least to the work of Von Neumann \cite{vN}, and include also attempts by Schwinger \cite{Schwinger}, Mackey \cite{Mackey}, Ludwig \cite{Ludwig}, Piron \cite{Piron}, and many others. However, the consensus is that these were not entirely successful: partly because the results they achieved (e.g., Piron's well-known 
representation theorem) did not rule out certain rather exotic alternatives to QM, but mostly because the axioms deployed seem, in retrospect, to lack sufficient physical or operational motivation. 
`
More recently, with inspiration from quantum information theory, attention has focussed 
on finite-dimensional systems, where the going is a bit easier. At the same time, and perhaps more 
significantly, quantum information theory prompts us to treat properties of composite systems as fundamental, where earlier work focussed largely on systems in isolation.\footnote{A recent exception to this trend is the paper 
\cite{BMU} of Barnum, M\"{u}ller and Ududec.}
These shifts of emphasis were strikingly illustrated by the work of Hardy \cite{Hardy}, 
who presented five simple, broadly information-theoretic postulates governing the states and measurements associated with a physical system, and 
showed that these lead to a very restricted class of theories parametrized by a positive integer, with  
 finite-dimensional quantum and classical probability theory as the first two entries.  
Following this lead, several recent papers, notably 
\cite{CDP, Dakic-Brukner, Masanes-Mueller}, have derived standard formulation of finite-dimensional QM  
from various packages of axioms governing the information-carrying and information-processing capacity of finite-dimensional probabilistic systems.  


\noindent{\bf Problems with existing approaches} These recent reconstructive efforts suffer from two related problems. First, they make use of assumptions that seem 
{\em too strong}; secondly, in trying to derive {\em exactly} complex, finite-dimensional quantum theory, they derive {\em too much}. 

All of the cited papers assume {\em local tomography}. This is the doctrine that the state of a bipartite composite system is entirely determined by the joint probabilities it assigns to outcomes of measurements on the two subsystems. This rules out both real and quaternionic QM, both of which are legitimate quantum theories \cite{Baez}. 
These papers also make some version of a {\em uniformity assumption}: that all systems having the same information-carrying capacity are isomorphic, or that all systems are composed, in a uniform way, from ``bits" of a uniform structure.\footnote{Here, ``information capacity" means essentially the maximum number of states that 
can be sharply distinguished (distinguished with probability one) from one another by a single measurement, 
and a bit is a system for which this number is two.} This rules out systems involving superselection rules, i.e., those that admit both real {\em and} classical degrees of freedom. \footnote{For example, the quantum system corresponding to $M_2(\C) \oplus M_2(\C)$,  corresponding to a classical 
choice between one of two qubits, has the same information-carrying capacity as a single, four-level quantum system.}
More seriously, it 
rules out any theory that includes, e.g., real {\em and} complex, or real {\em and} quaternionic systems, as the  state spaces of the bits of these theories have different dimensions. As I'll 
discuss below, one can indeed construct mathematically reasonable theories that embrace 
finite-dimensional quantum systems of all three types. 
An additional shortcoming, not related to the exclusion of real and quaternionic QM, is the assumption (explicit in \cite{Masanes-Mueller} for bits) that all positive affine functionals on the state space that take values 
between $0$ and $1$, represent physically accessible effects.  From an operational point of view, this principle (called the ``no-restriction hypothesis" in  \cite{Janotta-Lal}) 
seems to call for further motivation. 

\noindent{\bf Another approach} 
In these notes, I'm going to describe an alternative approach 
that avoids these difficulties. 
This begins by isolating  two striking features shared by classical and quantum probabilistic systems. 
The first is the possibility of finding a joint state that perfectly correlates a system $A$ with an isomorphic system $\bar{A}$ --- call it 
a  {\em conjugate} system --- in the sense that every basic observable on $A$ is perfectly correlated with the corresponding observable on $\bar{A}$. 
In finite-dimensional QM, where $A$ is represented by a 
finite-dimensional Hilbert space $\H$, $\bar{A}$ corresponds to the conjugate Hilbert space $\bar{\H}$, 
and the perfectly correlating state is the maximally entangled ``EPR" state on $\H \otimes \bar{\H}$. 

The second feature is the existence of what I call {\em filters} associated with basic observables. These are processes that independently attenuate the ``response" of each outcome by some specified factor.                 
Such a process will generally not preserve the normalization of states, but up to a constant  
factor, in both classical and quantum theory one can prepare any desired state by applying a 
suitable filter to the maximally mixed 
state. Moreover, when the target state is not singular (that is, when it does not 
assign probability zero to any nonzero measurement outcome), one can reverse the filtering process, in the sense that  
it can be undone by another process with positive probability. 

The upshot is that all probabilistic systems having conjugates and a 
sufficiently lavish supply of reversible filters can be represented by 
 {\em formally real Jordan algebras}, a class of structures that includes real, complex and quaternionic quantum systems, and just two further (and well studied) additional possibilities, which I'll review below.

In addition to leaving room for real and quaternionic quantum mechanics --- which I take to be a virtue --- 
this approach  has another advantage: {\em it is much easier!} The assumptions involved are few and 
easily stated, and the proof of the main technical result (Lemma 1 in Section 4) is short and straightforward. 
By contrast, the mathematical developments in the papers listed above are significantly more 
difficult, and ultimately lean on the (even more difficult) classification of compact groups acting on spheres. My approach, too, leans on a received result, but one that's relatively accessible. This is the Koecher-Vinberg theorem, which characterizes formally real, or euclidean, Jordan algebras in terms of ordered real vector spaces with homogeneous, self-dual cones. A short and non-taxing proof of this classical result can be found in \cite{FK}. 

These ideas were developed in \cite{Wilce09, Wilce11} and especially 
\cite{Wilce12}, of which this paper is, to an extent, a summary. 
However, the presentation here is slightly different,  and some additional ideas are also explored. I also briefly discuss recent work with Howard Barnum and Matthew Graydon \cite{BGraW} 
on the construction of probabilistic theories 
in which real, complex and quaternionic quantum systems 
coexist.

\noindent{\bf A bit of background} At this point, I'd better pause to  explain some terms. A {\em Jordan algebra} is a real commutative algebra --- a real vector space $\Jor$ with a commutative bilinear multiplication $a,b \mapsto a \dot b$ --- 
having a multplicative unit $u$, and satisfying the {\em Jordan identity}: 
$a^2 \dot (a \dot b) = a \dot (a^2 \dot b)$,
for all $a, b, c \in \Jor$, where $a^2 = a \dot a$.  A Jordan algebra is {\em formally real} if sums of squares of nonzero elements are always nonzero. The basic, and motivating, example is the space $\Ll(\H)$ of self-adjoint operators on a complex Hilbert space, with Jordan product given by $a \dot b = \frac{1}{2}(ab + ba)$. Note that here $a \dot a = aa$, so the 
notation $a^2$ is unambiguous. To see that $\Ll(\H)$ is formally real, just note that $a^2$ is always a positive 
operator. 

If $\H$ is finite dimensional, $\Ll(\H)$ carries a natural 
inner product, namely $\langle a, b \rangle = \Tr(ab)$. This plays well with the Jordan product: 
$\langle a \dot b, c \rangle = \langle b, a \dot c \rangle$. More generally, a finite-dimensional 
Jordan algebra equipped with an inner product having this property is said to be {\em euclidean}. 
For finite-dimensional Jordan algebras,  
being formally real and being euclidean are equivalent \cite{FK}. In what follows, I'll abbreviate ``euclidean 
Jordan algebra" to EJA. 

Jordan algebras were originally proposed, with what now looks like slightly thin motivation, by P. Jordan \cite{Jordan}: if $a$ and $b$ are quantum-mechanical observables, represented by $a, b \in \Ll(\H)$, then while $a + b$ is again self-adjoint, $ab$ is not, unless $a$ and $b$ commute; however, their average, $a \dot b$, {\em is} self-adjoint, and thus, represents another observable. 
Almost immediately, Jordan, von Neumann and Wigner showed \cite{JNW}
that all formally real Jordan algebras 
are direct sums of simple such algebras, with the 
latter falling into just five classes, parametrized by positive integers $n$: self-adjoint parts 
of matrix algebras $M_n(\F)$, where $\F = \R, \C$ or $\Q$ (the quaternions) or, for $n = 3$, 
over $\Oct$ (the Octonions); and also what are called {\em spin factors} $V_n$ (closely related to Clifford algebras).
\footnote{There is some overlap: $V_2 \simeq M_2(\R), V_3 \simeq M_2(\C)$ and $V_5 \simeq M_2(\Q)$.}
In all but one case, one can show that a simple Jordan algebra is a Jordan subalgebra of $M_n(\C)$ for 
suitable $n$. The {\em exceptional Jordan algebra}, $M_3(\Oct)_{\sa}$, admits no such representation. 

Besides this classification theorem, there is only one other important fact about euclidean Jordan 
algebras that's needed for what follows. This is the Koecher-Vinberg (KV) theorem alluded to above. 
Any EJA is also an ordered vector space, with positive cone $\Jor_+ := \{ a^2 | a \in A\}$.\footnote{Recall here than an ordered vector space is a real vector space, call 
it $\Jor$, spanned by a distinguished convex cone $\Jor_+$ having its vertex at the origin. Such a cone 
induces a translation-invariant partial order on $\Jor$, namely $a \leq b$ iff $b - a \in \Jor_+$.} This cone has two 
special features: first, it is {\em homogeneous}, i.e, for any points $a, b$ in the {\em interior} of 
$\Jor_+$, there exists an automorphism of 
the cone --- a linear isomorphism $\Jor \rightarrow \Jor$, taking $\Jor_+$ onto itself --- that maps $a$ onto $b$. 
In other words, the group of automorphisms of the cone acts transitively on the cone's interior. 
The other special property is that $\Jor_+$ is {\em self-dual}. This means that $\Jor$ carries an inner product 
--- in fact, the given one making $\Jor$ euclidean --- such that $a \in \Jor_+$ iff $\langle a, b \rangle \geq 0$ 
for all $b \in \Jor_+$. In the following, by a {\em euclidean order unit space}, I mean an ordered vector 
space $\Jor$ equipped with an inner product $\langle, \rangle$ with $\langle a, b \rangle \geq 0$ 
for all $a, b \in \Jor_+$, and a distinguished order-unit $u$. I will say that such a space $\Jor$ is 
{\em HSD} iff $\Jor_+$ is homogeneous, and also self-dual with respect to the given inner prouct. 

\tempout{The simplest example is the space $\R^{n}$, with its usual inner product, ordered by the {\em positive orthant}, i.e, 
the set of vectors having non-negative entries. A more interesting example is the space $\L(\H)$ of 
self-adjoint operators on a finite-dimensional Hilbert space, with the trace inner product, ordered in 
the usual way (namely, $a \leq b$ iff $\langle a x, x \rangle \leq \langle b x, x \rangle$ for all $x \in \H$). 
More generally, one can show that any euclidean Jordan algebra $\Jor$, ordered by the cone $\E_{+} = \{ a^2 | a \in \E\}$, is both homogeneous and self-dual. 
}

\noindent{\bf Theorem [Koecher 1958; Vinberg 1961]:} {\em Let $\E$ be a finite-dimensional euclidean order-unit 
space. If $\E$ is HSD, then there exists a unique product $\dot$ with respect to which $\E$ (with its given inner product) is a 
euclidean Jordan algebra, $u$ is the Jordan unit, and $\E_+$ is the cone of squares.} 

It seems, then, that if we can motivate a representation of physical systems in terms of HSD order-unit spaces, 
we will have ``reconstructed" what we might call (with a little license) finite-dimensional {\em Jordan}-quantum mechanics. In view of the classification theorem glossed above, this gets us into the neighborhood of orthodox QM, but still leaves open the possibilty of taking real and quaternionic quantum systems seriously. (It also leaves the 
door open to two possibly unwanted guests, namely spin factors and the exceptional Jordan algebra. I'll discuss 
below some constraints that at least bar the latter.)


\tempout{Very general (and pretty well known) considerations show that the any a probabilistic 
system --- one characterized by states and observables, with states assigning probabilities to outcomes of 
observables --- can be modeled in terms of an order-unit space. In view of the Koecher-Vinberg Theorem, one 
would like to find a set (ideally, a small set!) of simple, plausible, physically (or probabilistically, or operationally) meaningful assumptions, from which it follows that this ordered vector space is HSD. 
In what follows, I'll present, not just one, but two such packages (one involving four, the other, just two 
separate assumptions.)}



{\bf Some Notational Conventions}: My notation is {\em mostly} consistent with the following conventions (more 
standard in the mathematics than the physics literature, but in places slightly excentric relative to either). Capital Roman letters $A, B, C$ serve as labels 
for systems. Vectors in a Hilbert space $\H$ are denoted by little roman letters $x,y,z$ from the end of the alphabet.                                            Operators on $\H$ will usually be denoted by little roman letters $a,b,c,...$ from the begining of the alphabet. 
Roman letters $t, s$ typically stand for real numbers. The space of linear operators on $\H$ is denoted 
$\Ll(\H)$; $\Ll_{\sa}(\H)$ is the (real) vector space of self-adjoint operators on $\H$. 

As above, the conjugate Hilbert space is denoted $\bar{\H}$. 
I'll write $\bar{x}$ for the vectors in $\bar{\H}$ corresponding to $x \in \H$. From a certain point of view, 
this is the same vector; the bar serves to remind us that $\bar{cx} = \bar{c}~\bar{x}$ for scalars $c \in \C$. 
(Alternatively, one can regard $\bar{\H}$ as the space of ``bra" vectors $\langle x |$ corresponding to 
the ``kets" $|x \rangle$ in $\H$, i.e., as the dual space of $\H$.)

The inner product of $x, y \in \H$ is written as $\langle x,y \rangle$, and is linear in 
the {\em first} argument (if you like: $\langle x, y \rangle = \langle y | x \rangle$ in Dirac notation.). The 
inner product on $\bar{\H}$ is then $\langle \bar{x}, \bar{y} \rangle = \langle y, x \rangle$. The rank-one projection operator associated with a unit vector $x \in \H$ is $p_x$. 
Thus, $p_x(y) = \langle y, x \rangle x$.
I denote functionals on $\Ll_{\sa}(\H)$ by little Greek letters, e.g., $\alpha, \beta...$, and 
operators on $\Ll_{\sa}(\H)$ by capital Greek letters, e.g., $\Phi$. Two exceptions to this scheme: a generic density operator on $\H$ is denoted by the capital Roman letter $W$, and a certain unit vector in $\H \otimes \bar{\H}$ is denoted by the capital Greek letter $\Psi$. With luck, context will help keep things straight.

\section{Homogeneity and self-duality in quantum theory} 

Why {\em should} a probabilistic physical system be represented by a euclidean order-unit space that's either homogeneous or self-dual? 
One place to start hunting for an answer might be to look a standard quantum 
probability theory, to see if we can isolate, {\em in operational terms}, what makes {\em this} self-dual and homogeneous. 

\tempout{
Let $\H$ be a finite-dimensional complex Hilbert space, representing some finite-dimensional quantum system. As indicated above, 
$\L_{\sa}(\H)$ is the space of self-adjoint operators on $\H$. A state of the corresponding system is represented by 
a density operator (a positive operator of unit trace) $w \in \Ll_{\sa}(\H)$; a possible measurement outcome is represented 
by an {\em effect}, i.e., a positive operator $a \leq \1$. The probability of observing $a$ in the state corresponding 
to $w$ is $\tr(wa)$. If $w$ is the rank-one projection operator $p_{\phi}$ where $\phi$ is a unit vector in $\H$, 
then this is $\langle a \phi, \phi \rangle$. }

{\bf Correlation and self-duality} Let $\H$ be a finite-dimensional complex Hilbert space, representing some finite-dimensional quantum system. As above, $\Ll_{\sa}(\H)$ is the space of self-adjoint operators on $\H$. The system's states are represented by density operators, i.e., positive trace-one operators $W \in \Ll_{\sa} (\H)$; possible measurement-outcomes are 
represented by effects, i.e., positive operators $a \in \Ll_{\sa}(\H)$ with $a \leq \1$. The probability of 
observing effect $a$ in state $W$ is $\Tr(Wa)$. If $W$ is a pure state, i.e., $W = p_{v}$ where $v$ is a unit vector in $\H$, then $\Tr(Wa) = \langle a v, v \rangle$; by the same token, if $a = p_x$, then $\Tr(Wa) = \langle W x, x \rangle$. 

For $a, b \in \Ll_{\sa}(\H)$, let $\langle a, b \rangle = \Tr(ab)$. This is an inner product. By the spectral theorem, 
$\Tr(ab) \geq 0$ for all $b \in \Ll_{h}(\H)_+$ iff $\Tr(a p_x) \geq 0$ for all unit vectors $x$. But $\Tr(a p_x) = \langle a x, x \rangle$. So $\Tr(ab) \geq 0$ for all $b \in \Ll_{h}(\H)_+$ iff $a \in \Ll_{h}(\H)_+$, i.e., the trace inner product is self-dualizing. 
But this now leaves us with the 

{\bf Question:} {\em What does the trace inner product {\em represent}, probabilistically?}

Let $\bar{\H}$ be the conjugate Hilbert space to $\H$. Suppose $\H$ has dimension $n$. Any unit vector $\Psi$ 
in $\H \otimes \bar{\H}$ gives rise to a joint probability assignment to effects $a$ on $\H$ and $\bar{b}$ on $\bar{\H}$, namely $\langle (a \otimes \bar{b}) \Psi, \Psi \rangle$. Consider the 
{\em EPR state} for $\H \otimes \bar{\H}$ defined by the unit vector 
\[\Psi = \frac{1}{\sqrt{n}} \sum_{x \in E} x \otimes \bar{x} \in \H \otimes \bar{\H},\]
where $E$ is any orthonormal basis for $\H$. 
A straightforward computation shows that  
\[\langle (a \otimes \bar{b})\Psi, \Psi \rangle = \tfrac{1}{n} \Tr(ab).\]
In other words, {\em the normalized trace inner product {\em just is} the joint probability function $\eta$ determined by the pure state vector $\Psi$}! 

As a consequence, the state represented by $\Psi$ has a very strong correlational property: if $x, y$ are two orthogonal unit vectors with corresponding rank-one projections $p_x$ and $p_y$, we 
have $p_x p_y = 0$, so 
$\langle (p_x \otimes \bar{p_y}) \Psi, \Psi \rangle = 0$. 
On the other hand, 
$\langle (p_x \otimes \bar{p_x}) \Psi, \Psi \rangle = \tfrac{1}{n} \Tr(p_x) = \frac{1}{n}$. 
Hence, $\eta$ {\em perfectly, and uniformly, correlates} every basic measurement (orthonormal basis) of $\H$ with its counterpart in $\bar{\H}$. 

{\bf Filters and homogeneity} {\red Next, let's see why the cone $\Ll_{h}(\H)_+$ is homogeneous. Recall that this means that any 
state in the interior of the cone --- here, any non-singular density operator --- can be obtained from any other by an order-automorphism 
of the cone.  But in fact, something better is true: this order-automorphism can 
be chosen to represent a probabilistically reversible physical process, i.e., an invertible CP mapping with CP inverse.}

To see how this works, suppose $W$ is a positive operator on $\H$. Consider the pure CP mapping $\Phi_{W} : \Ll_{\sa}(\H) \rightarrow \Ll_{\sa}(\H)$ given by 
\[\Phi_{W}(a) = W^{1/2} a W^{1/2}.\]
Then $\Phi_{W}(\1) = W$. If $W$ is nonsingular, so is $W^{1/2}$, so $\Phi_{W}$ is invertible, with inverse 
$\Phi_{W}^{-1} = \Phi_{W^{-1}}$, again a pure CP mapping. Now given another nonsingular density operator $M$, 
we can get from $W$ to $M$ by applying $\Phi_{M} \circ \Phi_{W^{-1}}$. 

All well and good, but we are still left with the 

{\bf Question:} {\em What does the mapping $\Phi_{W}$ {\em represent}, physically?}

To answer this, suppose $W$ is a density operator, with spectral expansion $W = \sum_{x \in E} t_x p_x$. Here, $E$ is an orthonormal basis for $\H$ diagonalizing $W$, and $t_x$ is the eigenvalue of $W$ corresponding to $x \in E$.
Then, for each vector $x \in E$,  
\[\Phi_{W}(p_x) = t_x p_x\]
where $p_x$ is the projection operator associated with $x$. 
We can understand this to mean that $\Phi_W$ acts as a {\em filter} on the test $E$: the {\em response} of each outcome $x \in E$ is attenuated by a factor $0 \leq t_x \leq 1$.\footnote{My usage here is slightly non-standard, in that I 
allow filters that ``pass" the system with a probability strictly between 0 and 1.} Thus, if $M$ is another density operator on $\H$, representing some state of the corresponding system, then the probability of obtaining outcome $x$ after preparing the system in state $M$ and applying the process $\Phi$, is $t_x$ times the probability of $x$ in state $M$. (In detail: suppose $p_x$ is the rank-one projection operator 
associated with $x$, and note that $W^{1/2} p_x = p_x W^{1/2} = t_{x}^{1/2} p_x$. Thus, 
\begin{eqnarray*}
\Phi_{W}(M)(x)  = \Tr(W^{1/2} M W^{1/2} p_x) & = & \Tr(W^{1/2} M t_{x}^{1/2} p_x) = \Tr(t_{x}^{1/2} p_x W^{1/2} M) \\
& = & \Tr(t_{x} p_x M) = t_{x} \Tr(M p_x).
\end{eqnarray*}
If we think of the basis $E$ as representing a set of alternative {\em channels} plus {\em detectors}, as in the figures 
below, we can add a classical filter attenuating the response of one of the detectors --- say, $x$ --- by a fraction $t_x$  What the computation above tells us is that we can achieve the same result by applying a suitable 
CP map to the system's state. Moreover, this can be done independently for each outcome of $E$. In Figure 1 below, 
this is illustrated for 3-level quantum system: $E = \{x,y,z\}$ is an orthonormal basis, representing three 
possible outcomes of a Stern-Gerlach-like experiment; the filter $\Phi$ acts on the system's state in such a 
way that the probability of outcome $x$ is attenuated by a factor of $t_x = 1/2$, while outcomes $y$ and $z$ are 
unaffected.  
\begin{center}
\begin{tikzpicture} 
\node[anchor=east] at (0,0) (source) {$\alpha$};
\node[anchor=west] at (2,0) (splitter) {$\stackrel{\bigtriangledown}{\triangle}$}; 
\node[anchor=west] at (5,1) (detectorx) {$x$}; 
\node[red] at (7,1) {prob = $\frac{1}{2} \alpha(x)$};
\node[anchor=west] at (5,0) (detectory) {$y$};
\node at (7,0) {prob = $ \alpha(y)$};
\node[anchor=west] at (5,-1) (detectorz) {$z$};
\node at (7,-1) {prob = $\alpha(z)$};
\draw (source) edge[out=0,in=180,->] (splitter);
\draw (splitter) edge[out=0,in=180,->] (detectorx);
\draw (splitter) edge[out=0,in=180,->] (detectory);
\draw (splitter) edge[out=0,in=180,->] (detectorz);
\node at (1,0) (ring) {};
\draw[red] (1,0) circle(.2cm); 
\node[red] at (.5,-1) (filterlabel) {$\Phi$};
\draw[red] (filterlabel) edge[out=45,in=225,->] (ring);
\end{tikzpicture}\\
\vspace{.1in}
{Figure 1: $\Phi$ attenuates $x$'s sensitivity by $1/2$. }
\end{center}
\tempout{
\begin{center}
\begin{tikzpicture}
\draw[->] (0,0) -- (2,0) -- (3,0); 
\draw[->] (2,0) -- (2.2,1) -- (3,1);
\draw[->] (2,0) -- (2.2,-1) -- (3,-1); 
\draw (3.2,1) node{$x$};
\draw (3.2,0) node{$y$};  
\draw (3.2,-1) node{$z$};
\draw (2.7,1) node{$\Box$};
\draw (0,-.2) node{$\alpha$};
\end{tikzpicture} 
{Figure (a)}
\hspace{.2in}
\begin{tikzpicture}
\draw[->] (0,0) -- (2,0) -- (3,0); 
\draw[->] (2,0) -- (2.2,1) -- (3,1);
\draw[->] (2,0) -- (2.2,-1) -- (3,-1); 
\draw (3.2,1) node{$x$};
\draw (3.2,0) node{$y$};  
\draw (3.2,-1) node{$z$};
\draw (1.5,0) node{$\Box$};
\draw (1.5,.3) node{$\Phi_x$};
\draw (0,-.2) node{$\alpha$};
\end{tikzpicture}
{Figure (b)}
\end{center} 
}
If we apply the filter $\Phi_{W}$ to the maximally mixed state $\tfrac{1}{n} \1$, we obtain $\tfrac{1}{n} W$. Thus, we can {\em prepare} $W$, up to normalization, by applying the filter $\Phi_W$ to the maximally mixed state. 

{\bf Filters are Symmetric} Here is a final observation, linking these last two: The filter $\Phi_{W}$ is {\em symmetric} with respect to the uniformly correlating state $\eta$, in the sense that 
\[\langle (\Phi_{W}(a) \otimes \bar{b}) \Psi, \Psi \rangle = \langle (a \otimes \bar{\Phi}_{W}(b))\Psi, \Psi \rangle \]
for all effects $a, b \in \Ll_{\sa}(\H)_{+}$. 
Remarkably, {\em this is all that's needed} to recover the Jordan structure of finite-dimensional quantum theory: the existence of a conjugate system, with a uniformly 
correlating joint state, plus the possibility of preparing non-singular states by means of filters that are symmetric with 
respect to this state, and doing so reversibly when the state is nonsingular. 

In very rough outline, the 
argument is that states preparable (up to normalization) by symmetric filters have spectral decompositions, and the existence of spectral decompositions makes the uniformly correlating joint state a self-dualizing inner product.  But to spell this out in any precise way, I need a general mathematical framework for discussing states, effects and processes in abstraction from quantum theory. The next section reviews the necessary apparatus.

\section{General probabilistic theories} 

A characteristic feature of quantum mechanics is the existence of incompatible, or non-comeasurable, observables. This suggests the following simple, but very fruitful, notion: 

{\bf Definition:} A {\em test space} is a collection $\M$ of non-empty sets $E, F, ....$, each representing the {\em outcome-set} of some measurement, experiment, or {\em test}. At the 
outset, one makes no special assumptions about the combinatorial structure of $\M$. In particular, distinct tests are permitted to overlap.  Let $X := \bigcup \M$ denote the set of all 
outcomes of all tests in $\M$: a {\em probability weight} on $\M$ is a function $\alpha : X \rightarrow [0,1]$ such that $\sum_{x \in E} \alpha(x) = 1$ for every $E \in \M$.\footnote{Test spaces were introduced and studied by D. J. Foulis and C. H. Randall and their students (of whom I'm one) in a long series of papers beginning around 1970. The original term for a test was an {\em operation}, which has the advantage of signaling that the concept has wider applicability than simply reading a number off a meter: anything an agent can {\em do} that leads to a well-defined, exhaustive set of mutually exclusive outcomes, defines an operation. Accordingly, test spaces were originally called ``manuals of operations".}

It can happen that a test space admits no probability weights at all. However, to serve as a model of a real family of experiments associated with an actual physical system, a test space should obviously carry a lavish supply of such weights. One might want to single out some of these as describing physically (or otherwise) possible states of the system.   This suggests the following 

{\bf Definition:} A {\em probabilistic model} is a pair $A = (\M,\Omega)$, where $\M$ is a test space and $\Omega$ is some designated convex set of probability weights, called 
the {\em states} of the model. 

The definition is deliberately spare. Nothing prohibits us from adding further structure (a group of symmetries, say, or a toplogy on the space of outcomes). However, no such additional structure is needed for the results I'll discuss below.   I'll write $\M(A), X(A)$ and $\Omega(A)$ for the tet space, associated outcome space, and state space of a model $A$. The convexity asumption on $\Omega(A)$ is intended to capture the possibility of forming mixtures of states. To allow the modest idealization of taking outcome-wise limits of states to be states, 
I will also assume that $\Omega(A)$ is 
closed as a subset of $[0,1]^{X(A)}$ (in its product topology). Note that this makes $\Omega(A)$ compact, and so, guaranteees the existence of pure states, that is, extreme points of $\Omega(A)$. If $\Omega(A)$ is the set of {\em all} probability weights on $\M(A)$, I'll say that $A$ has a {\em full state space}. 

{\bf Two Bits}   Here is a simple but instructive illustration of these notions. Consider a test space $\M = \{\{x,x'\}, \{y, y'\}\}$. Here we have two tests, each with two outcomes. We are permitted to perform either test, but not both at once. A probability weight is determined by the values it assigns to $x$ and to $y$, and --- since the sets $\{x,x'\}$ and $\{y,y'\}$ are disjoint --- these values are independent. Thus, geometrically, the space of all probability weights is the unit square  in $\R^2$ (Figure 2(a), below). To construct a probabilistic model, we can choose any closed, convex subset of the square for $\Omega$. For instance, we might let $\Omega$ be the convex hull of the four probability weights $\delta_x, \delta_{x'}$, $\delta_{y}$ 
and $\delta_{y'}$ corresponding to the midpoints of the four sides of the square, as in Figure 2(b) --- that is, 
\[\delta_{x}(x) = 1, \ \delta_{x}(x') = 0, \  \delta_{x}(y) = \delta_{x}(y') = 1/2,\]
\[\delta_{x'}(x) = 0, \ \delta_{x'}(x') = 1, \  \delta_{x'}(y) = \delta_{x'}(y') = 1/2,\]
and similarly for $\delta_{y}$ and $\delta_{y'}$. 
\[
\begin{array}{ccc}
\mbox{
\begin{tikzpicture} 
\draw[->] (0,0) -- (2.3,0); 
\draw[->] (0,0) -- (0,2.3);
\draw (2.5,0) node{$x$}  (0,2.5) node{$y$};
\draw[-] (2,0) -- (2,-.2);
\draw[-] (0,2) -- (-.2,2);
\draw (-.4,2) node{$1$}  (2,-.4) node{$1$}; 
\draw[-, fill=gray] (0,0) -- (2,0) -- (2,2) -- (0,2) -- (0,0); 
\end{tikzpicture}
}
& & 
\mbox{
\begin{tikzpicture} 
\draw[-] (0,0) -- (2,0) -- (2,2) -- (0,2) -- (0,0); 
\draw[-, fill=gray] (1,0) -- (2,1) -- (1,2) -- (0,1) -- (1,0);
\draw (1,-.25) node{$\delta_{y'}$};
\draw (2.2,1) node{$\delta_{x}$};  
\draw (1,2.2) node{$\delta_{y}$};
\draw (-.25,1) node{$\delta_{x'}$};
\end{tikzpicture}
}\\
\mbox{Figure 2(a)} & & \mbox{Figure 2(b)} 
\end{array} 
\]
The model of Figure 2(a), in which we take $\Omega$ to the be entire set of probability weights on $\M = \{\{x,x'\}, \{y,y'\}\}$ is sometimes called the {\em square bit}. I'll call the model of Figure 2(b) the {\em diamond bit}.

{\bf Classical, Quantum and Jordan Models} If $E$ is a finite set, the corresponding {\em classical model} is $A(E) = (\{E\}, \Delta(E))$ where $\Delta(E)$ is the simplex of probability weights on $E$. If $\H$ is a finite-dimensional complex Hilbert space, let $\M(\H)$ denote the set of orthonormal bases of $\H$: then $X = \bigcup \M(\H)$ is the unit sphere of 
$\H$, and any density operator $W$ on $\H$ defines a probability weight $\alpha_W$, given by $\alpha_W(x) = \langle W x, x \rangle$ for all $x \in X$. Letting 
$\Omega(\H)$ denote the set of states of this form, we obtain the {\em quantum model}, $A(\H) = (\M(\H), \Omega(\H))$, associated with $\H$.\footnote{The content of Gleason's Theorem 
is that $A(\H)$ has a full state space for $\dim(\H) > 2$.  We will not need this fact.}

More generally, every euclidean Jordan algebra $\Jor$ gives rise to a probabilistic model as follows. A minimal or {\em primitive} idempotent of $\Jor$ is an element $p \in \Jor$ with 
$p^2 = p$ and, for $q = q^2 < p$, $q = 0$. A {\em Jordan frame} is a maximal pairwise orthogonal set of primitive idempotents. Let $X(\Jor)$ be the set of primitive idempotents, 
let $\M(\Jor)$ be the set of Jordan frames, and let $\Omega(\Jor)$ be the set of probability weights of the form $\alpha(p) = \langle a, p \rangle$ where $a \in \Jor_+$ with 
$\langle a, u \rangle = 1$. This data defines the {\em Jordan model} $A(\Jor)$ associated with $\Jor$. In the case where $\Jor = \Ll_{h}(\H)$ for a finite-dimensional Hilbert space $\H$, 
this {\em almost} gives us back the quantum model $A(\H)$: the difference is that we replace unit vectors by their associated projection operators, thus conflating outcomes that differ only by a phase. 

\noindent{\bf Sharp Models} Jordan models enjoy many special features that the generic probabilistic model lacks. I want to take a moment to disuss one such feature, which will be important below. 

{\bf Definition:}  A model $A$ is {\em unital} iff, for every outcome $x \in X(A)$, there exists a state $\alpha \in \Omega(A)$ with $\alpha(x) = 1$, and {\em sharp} if this state is 
unique (from which it follows easily that it must be pure). If $A$ is sharp, I'll write $\delta_x$ for the unique state making $x \in X(A)$ certain. 

If $A$ is sharp, then there is a sense in which each test $E \in \M(A)$ is {\em maximally informative:} if we are certain which outcome $x \in E$ will occur,  
then we know the system's state exaclty, as there is only one state in which $x$ has probability  $1$. 

Classical and quantum models are obviously sharp. More generally, every Jordan model is sharp. To see this, note first that every state $\alpha$ on a euclidean Jordan algebra $\Jor$ 
has the form $\alpha(x) = \langle a, x \rangle$ where $a \in \Jor_+$ with $\langle a, u \rangle = 1$, and where $\langle \ , \ \rangle$ is the given inner product on $\Jor$, normalized 
so that $\|x\| = 1$ for all primitive idempotents (equivalently, so that $\|u\| = n$, the rank of $\Jor$). The spectral theorem for EJAs shows that 
$a= \sum_{p \in E} t_p p$ where $E$ is a Jordan frame and the coefficients $t_p$ are non-negative and sum to $1$ (since $\langle a, u \rangle = 1$) . If 
$\langle a, x \rangle = 1$, then $\sum_{p \in E} t_p \langle p, x \rangle = 1$ implies that, for every $p \in E$ with $t_p > 0$, $\langle p, x \rangle = 1$. But 
$\|p\| = \|x\| = 1$, so this implies that $\langle p, x \rangle = \|p\|\|x\|$ which in turn implies that $p = x$.

In general, a probabilistic model need not even be unital, much less sharp. On the other hand, given a unital model $A$, it is often possible to construct a sharp model by suitably restricting the state space. This is illustrated in Figure 2(b) above: the full state space of the square bit is unital, but far from sharp; however, by restricting the state space to the convex hull of the barycenters of the faces, we obtain a sharp model. This is possible  whenever $A$ is unital and carries a group of symmetries acting transitively on the outcome-set $X(A)$. For details, see Appendix A. The point here is that sharpness is not, by itself, a very stringent condition: since we should expect to encounter highly symmetric, unital models abundantly ``in nature", we can also expect to encounter an abundance of sharp models.

\noindent{\bf The spaces $\V(A)$, $\V^{\ast}(A)$}  Any probabilistic model gives rise to a pair of ordered vector spaces in a canonical way. These will be essential in the development below, so I'm going to go into a bit of detail here. 

{\bf Definition:} Let $A$ be any probabilistic model. Let $\V(A)$ be the span of the state space $\Omega(A)$ in $\R^{X(A)}$, ordered by the cone $\V(A)_+$ consisting of non-negative multiples of states, i.e., 
\[\V(A)_{+} = \{ t\alpha | \alpha \in \Omega(A), \ t \geq 0\}.\]
Call the model $A$ {\em finite-dimensional} iff $\V(A)$ is finite-dimensional. {\em From now on, I assume that all models are finite-dimensional.}

Let $\V^{\ast}(A)$ denote the dual space of $\V(A)$, ordered by the {\em dual cone} of positive linear functionals, i.e., functionals $f$ with $f(\alpha) \geq 0$ for all $\alpha \in \V(A)_{+}$. 
Any measurement-outcome $x \in X(A)$ yields an evaluation functional $\x \in \V^{\ast}(A)$, given by $\x(\alpha) = \alpha(x)$ for all $\alpha \in \V(A)$. More generally, 
an {\em effect} is a positive linear functional  $f \in \V^{\ast}(A)$ with $0 \leq f(\alpha) \leq 1$ for every state $\alpha \in \Omega(A)$.  The functionals $\x$ are effects. 
One can understand an arbitrary effect $a$ to represent a {\em mathematically} posssible measurement outcome, having probability $a(\alpha)$ in state $\alpha$. I 
stress the adjective {\em mathematically} because, {\em a priori}, there is no guarantee that every effect will correspond to a {\em physically} realizable measurement outcome. 
In fact, at this stage, I make no assumption at all about what, apart from the tests $E \in \M(A)$, is or is not physically realizable. (Later, it will follow from 
further assumptions that every element of $\V^{\ast}(A)$ represents a random variable associated with some $E \in \M(A)$, and is, therefore, operationally meaningful. But 
this will be  a theorem, not an assumption.) 

The {\em unit effect} is the functional $u_A := \sum_{x \in E} \x$, where $E$ is {\em any} element of $\M(A)$. This takes the constant value $1$ on $\Omega(A)$, and 
thus, represents a trivial measurement outcome that occurs with probability one in every state. This is an {\em order unit} for $\V^{\ast}(A)$: given any 
$a \in \V^{\ast}(A)$, one can find some constant $N > 0$ such that $a \leq Nu_A$. (To see this, just let $N$ be the maximum value of $|a(\alpha)|$ for $\alpha \in \Omega(A)$, 
remembering that the latter is compact.) 

For both classical and quantum models, the ordered vector spaces 
$\V^{\ast}(A)$ and $\V(A)$ are naturally isomorphic. If $A(E)$ is the classical model associated with a finite set $E$, both are isomorphic to 
the space $\R^{E}$ of all real-valued functions on $E$, ordered pointwise. 
If $A = A(\H)$ is the quantum model associated with a finite-dimensional Hilbert space $\H$, $\V(A)$ and $\V^{\ast}(A)$ 
are both naturally isomorphic to the space $\Ll_{h}(\H)$ of hermitian operators on $\H$, ordered by its usual cone of positive semi-definite operators. More generally, if $\Jor$ is a euclidean Jordan algebra and $A = A(\Jor)$ is the 
corresponding Jordan model, then $\V(A) \simeq \V^{\ast}(A) \simeq \Jor$, the latter ordered by its cone of squares. 

{\bf The space $\E(A)$} It's going to  be technically useful to introduce a third ordered vector space, which I will denote by $\E(A)$. This is the span of the evaluation-effects $\x$, associated with 
measurement outcomes $x \in X(A)$, in $\V^{\ast}(A)$, ordered by the cone 
\[\E(A)_+ \ := \ \left \{ \sum_{i} t_i \x_i | t_i \geq 0\right \}.\]
That is, $\E(A)_+$ is the set of linear combinations of effects $\x$ having non-negative coefficients. It is important to note that this is, in general, a {\em proper} sub-cone 
of $\V(A)_+$. To see this, revisit again example  1:  Here, $x$ is the outcome corresponding to the right face of the larger (full) state space pictured on the left, while 
$y$ is the outcome corresponding to the top face. Now consider the functional $f := \x + \y - \frac{1}{2}u$. This takes positive values on the smaller 
state space $\Omega$, but is negative on, for example, the state $\gamma$ corresponding to the lower-left corner of the full state space. Thus, $f \in \V(A)_{+}$, but 
$f \not \in \E(A)$.  
\[
\begin{array}{ccc}
\mbox{
\begin{tikzpicture} 
\draw[gray,fill=gray] (0,0) -- (2,0) -- (2,2) -- (0,2) -- (0,0); 
\draw[-, thick, red] (2,2.4) -- (2,-.4);
\draw (2.6,1) node{$\x = 1$};
\draw[ -, thick, red] (-0.4,2) -- (2.4,2);
\draw (1,2.25) node{$\y = 1$};
\draw[-,white] (0,-.43) node{};
\end{tikzpicture}
}
& & 
\mbox{
\begin{tikzpicture} 
\draw[-] (0,0) -- (2,0) -- (2,2) -- (0,2) -- (0,0); 
\draw[-, fill=gray] (1,0) -- (2,1) -- (1,2) -- (0,1) -- (1,0);
\draw[-,thick,blue] (-.5,1.5) -- (1.5,-.5);
\draw (2,-.27) node{$f = 0$};
\draw[-,thick,red] (.5,2.5) -- (2.5,.5);
\draw (2.77,.9) node{$f = 1$};
\end{tikzpicture}
}\\
\mbox{Figure 3(a)} & & \mbox{Figure 3(b)} 
\end{array} 
\]
Since we are working in finite dimensions, the outcome-effects $\x$ span $\V^{\ast}(A)$. Thus, {\em as vector spaces}, $\E(A)$ and $\V^{\ast}(A)$ are the same. However, as the diamond bit illustrates, 
they can have quite different positive cones, and thus, need not be isomorphc as {\em ordered} vector spaces.



\noindent{\bf Processes and subnormalized states} A {\em subnormalized state} of a model $A$ is 
an element $\alpha$ of $\V(A)_+$ with $u(\alpha) < 1$. These can be understood as states that allow a nonzero probability $1 - u(\alpha)$ of some generic ``failure" event, (e.g., the destruction of the system), represented by the $0$ functional in $\V^{\ast}(A)$. 

More generally, we may wish to regard two systems, represented by models $A$ and $B$, 
as the input to and output from some {\em process}, whether dynamical or purely information-theoretic, that 
has some probability to destroy the system or otherwise ``fail".  In general, 
such a process should be represented mathematically by an affine mapping $\Omega(A) \rightarrow \V(B)_{+}$, taking 
each normalized state $\alpha$ of $A$ to a possibly {\em sub-normalized} state $T(\alpha)$ of $B$. 
One can show that such a mapping extends uniquely to a positive linear 
mapping $T : \V(A) \rightarrow \V(B)$, so from now on, this is how I represent processes. 

Even if a process $T$ has a nonzero probability of failure, it may be possible to reverse its effect with nonzero 
probability. 

{\bf Definition:} A process $T : A \rightarrow B$ is {\em probabilistically reversible} iff there exists a process $S$ such that, for all $\alpha \in \Omega(A)$, $(S \circ T)(\alpha) = p \alpha$, where $p \in (0,1]$. 

This means that there is a probability $1-p$ of the composite process $S \circ T$ failing, but a probability 
$p$ that it will leave the system in its initial state. (Note that, since $S \circ T$ is linear, $p$ must be constant.)
Where $T$ preserves normalization, so that $T(\Omega(A)) \subseteq \Omega(B)$, $S$ can also be taken to be normalization-preserving, and will undo the result of $T$ with probability 1. This is the more usual meaning of ``reversible" in the literature. 

Given a process $T : \V(A) \rightarrow \V(B)$, there is a dual mapping $T^{\ast} : \V^{\ast}(B) \rightarrow \V^{\ast}(A)$, 
also positive, given by $T^{\ast}(b)(\alpha) = b(T(\alpha))$ for all $b \in \V^{\ast}(B)$ and $\alpha \in \V(A)$. The 
assumption that $T$ takes normalized states to subnormalized states is equivalent to the requirement that 
$T^{\ast}(u_B) \leq u_A$, that is, that $T^{\ast}$ maps effects to effects.  


{\em Remark:} Since we are attaching no special physical interpretation to the cone $\E_+$, it is important that we do {\em not} require physical processes $T : \V(A) \rightarrow \V(B)$ 
to have dual processes $T^{\ast}$ that map $\E_{+}(B)$ to $\E_{+}(A)$. That is, we do {\em not} require $T^{\ast}$ to be positive as a mapping $\E(B) \rightarrow \E(A)$.   



{\bf Joint probabilities and joint states} 
If $\M_1$ and $\M_2$ are two test spaces, with outcome-spaces $X_1$ and $X_2$, we can construct a space of {\em product tests} 
\[\M_1 \times \M_2 = \{ E \times F | E \in \M_1, F \in \M_2\}\footnote{Note here the 
savage abuse of notation: $\M_1 \times \M_2$ is {\em not} the Cartesian product of $\M_1$ and $\M_2$.} \]
This models a situation 
in which tests from $\M_1$ and from $\M_2$ can be performed separately, and the results colated.  Note that the outcome-space for $\M_1 \times \M_2$ is $X_1 \times X_2$. A {\em joint probability weight} on $\M_1$ and $\M_2$ is just a probability 
weight on $\M_1 \times \M_2$, that is, a function $\omega : X_1 \times X_2 \rightarrow [0,1]$ such that $\sum_{(x,y) \in E \times F} \omega(x,y) = 1$ for all tests $E \in \M_1$ and $F \in \M_2$.   One says that 
$\omega$ is {\em non-signaling} iff the {\em marginal} (or {\em reduced}) probability weights $\omega_1$ and $\omega_2$, given by  
\[\omega_1(x)  = \sum_{y \in F} \omega(x,y) \ \ \mbox{and} \ \ \omega_2(y) = \sum_{x \in E} \omega(x,y)\]
are well-defined, i.e., independent of the choice of the tests $E$ and $F$, respectively. One can understand this to mean that the choice of which test to measure on $\M_1$ has no observable, i.e., no statistical, influence on the outcome of tests made 
of $\M_2$, and vice versa.    In this case, one also has well-defined {\em conditional} probability weights 
\[\omega_{2|x}(y) := \omega(x,y)/\omega_1(x) \ \ \mbox{and} \ \ \omega_{1|y} := \omega(x,y)/\omega_{2}(y)\]  
(with, say, $\omega_{2|x} = 0$ if $\omega_{1}(x) = 0$, and similarly for $\omega_{1|y}$).  This gives us the following bipartite version of the law of total probability \cite{FR}: for any choice 
of $E \in \M_1$ or $F \in \M_2$, 
\begin{equation} \omega_2 = \sum_{x \in E} \omega_1(x) \omega_{2|x} \ \mbox{and} \ \ \omega_1 = \sum_{y \in F} \omega_2(y) \omega_{1|y}.\end{equation}

{\bf Definition:} A {\em joint state} on a pair of probabilistic models $A$ and $B$ is a non-signaling joint probability weight $\omega$ on $\M(A) \times \M(B)$ such that, for 
every $x \in X(A)$ and every $y \in X(B)$, the conditional probability weights $\omega_{2|x}$ and $\omega_{1|y}$ are valid states in $\Omega(A)$ and $\Omega(B)$, respectively. It follows from (1) that 
the marginal weights $\omega_1$ and $\omega_2$ are also states of $A$ and $B$, respectively.\footnote{Notice that 
I do not, at this point, offer any notion of a composite model for a joint state to be a state {\em of}. The major 
results of this section do not depend on such a notion. Composite models will be discussed later, in Section 6, where, 
among other things, I will assume that states of such models induce (but may not be determind by) joint states in the 
sense defined here.} 


This naturally suggests that one should define, for models $A$ and $B$, a {\em composite model} $AB$, the states of which would be precisely the joint states on $A$ and $B$. If one takes $\M(A B) = \M(A) \times \M(B)$, this is essentially the ``maximal tensor product" of $A$ and $B$ \cite{BarnumWilceFoils}.  However, this does not coincide 
with the usual composite of quantum-mechanical systems. In section 6, I will discuss composite systems in more detail. Meanwhile, for the main results of this paper, the idea of a joint state is sufficient.

For a simple example of a joint state that is neither classical nor quantum, let $B$ denote the ``square bit" model discussed above. That is, $B = (\boldsymbol{\mathcal B},\Omega)$ where 
e $\boldsymbol{\mathcal B} = \{\{x,y\}, \{a,b\}\}$ is a test space with two non-overlapping, two-outcome tests, and $\Omega$ is the set of all probability weights thereon, amounting to the 
unit square in $\R^{2}$. 
The joint state on $B \times B$ given by the table below (a variant of the ``non-signaling box" of
Popescu and Rohrlich \cite{PR}) is clearly non-signaling. Notice that it also establishes a perfect, uniform correlation between the outcomes of any test on the first system and its counterpart 
on the second.  
\[\begin{array}{|c|cc|cc|}
\hline
  &  x  &  y  &  a  &  b  \\
\hline
x & 1/2 &  0  & 1/2 &  0  \\ 
y & 0   & 1/2 &  0  & 1/2 \\ 
a & 0   & 1/2 & 1/2  & 0  \\ 
b & 1/2 &  0  &  0  & 1/2 \\
\hline
\end{array}\]

\noindent{\bf The conditioning map} 
     If $\omega$ is a joint state on $A$ and $B$, define the associated 
{\em conditioning maps} 
$\ \hat{\omega}: X(A) \rightarrow \V(B) \ \ \mbox{and} \ \ \hat{\omega}^{\ast} : X(B) \rightarrow \V(A)$ 
by 
\[\hat{\omega}(x)(y) = \omega(x,y)  = \hat{\omega}^{\ast}(y)(x)\]
for all $x \in X(A)$ and $y \in X(B)$. 
Note that $\hat{\omega}(x) = \omega_1(x) \omega_{2|x}$ for every $x \in X(A)$, i.e., 
$\hat{\omega}(x)$ can be understood as the {\em un-normalized} conditional state of $B$ given the outcome 
$x$ on $A$. Similarly, $\hat{\omega}^{\ast}(y)$ is the unnormalized conditional state of $A$ given 
outcome $y$ on $B$.

{\bf The conditioning map  linearized} The conditioning map $\hat{\omega}$ extends uniquely to a positive linear mapping 
$\E(A) \rightarrow \V(B)$, which I also denote by $\hat{\omega}$, such that $\hat{\omega}(\x) = \hat{\omega}(x)$ for all outcomes $x \in X(A)$. 
To see this, consider the linear mapping $T : \V^{\ast}(A) \rightarrow \R^{X(B)}$ 
defined, for $f \in \V^{\ast}(A)$, by $T(f)(y) = f(\hat{\omega}^{\ast}(y))$ for all  $y \in X(B)$. 
If $f = \x$,  we have $T(\x) = \omega_{1}(x) \omega_{2|x} \in \V(B)_{+}$, whence, 
for all $y \in X(B)$, 
$T(\x)(y) = \omega(x,y) = \hat{\omega}(x)(y)$.  
Since the evaluation functionals $\x$ span $\E(A)$, the range of $T$ lies in $\V(B)$, and, moreover, $T$ is {\em positive} on the cone $\E(A)_+$. 
Hence, as advertised, $T$ defines a positive linear mapping $\E(B) \rightarrow \V(A)$, extending $\hat{\omega}$. 
In the same way, $\hat{\omega}^{\ast}$ defines a positive linear mapping $\hat{\omega}^{\ast} : \E(B) \rightarrow \V(A)$. 

In general,  $\hat{\omega}$ need {\em not} take $\V^{\ast}(A)_+$ into $\V(B)_+$. This is the principal reason 
for working with $\E(A)$ rather than $\V^{\ast}(A)$.

\tempout{
{\red Notice that if $\omega$ is a nonsignaling state, then $\sum_{x \in E} \hat{\omega}  = \omega_{2} \in \V(B)$ is 
independent of $E \in \M(A)$. Conversely, suppose $\mu : X(A) \rightarrow \V(B)_+$ is a ``weight-valued weight", 
that is, $\sum_{x \in E} \mu(x)$ is independent of $E$. Then we have a linear 
mapping $\mu^{\ast} : \V^{\ast}(B) \rightarrow \R^{X(A)}$, which is easily seen to take values in 
the space $\V(\M(A))$ generate by the set of {\em all} probability weights on $A$. Thus, if $A$ has a full 
state space,  then $\mu = \hat{\omega}$ for a nonsignaling state 
on $A$ and $B$, namely, $\omega(x,y) = \mu(x)(y) = \mu^{\ast}(y)(x)$. }
}

\section{Conjugates and Filters} 

We are now in a position to abstract the two features of QM discussed earlier. 
Call a test space $(X,\M)$ {\em uniform} iff all tests $E \in \M$ have the 
same size. The test spaces associated with quantum models have this feature, and it is quite easy to generate many other 
examples (cf Appendix A).    
A uniform  test space, say with tests of size $n$, always admits at least one probability weight, namely, the {\em maximally mixed} probability weight $\rho(x) = 1/n$ for all $x \in X$.  I will say that a probabilistic {\em model} $A$ is uniform  if 
the test space $\M(A)$ is uniform  {\em and} the maximally mixed state $\rho$ belongs to $\Omega(A)$.

{\bf Definition:} Let $A$  be  uniform probabilistic model with tests of size $n$. 
A {\em conjugate} for $A$ is a model $\bar{A}$, 
plus a chosen isomorphism\footnote{By an {\em isomorphism} from a model $A$ to a model $B$, I mean the obvious thing: a bijection $\gamma : X(A) \rightarrow X(B)$ such 
taking $\M(A)$ onto $\M(B)$, and such that $\beta \mapsto \beta \circ \gamma$ maps $\Omega(B)$ onto $\Omega(A)$.}  $\gamma_A : A \simeq \bar{A}$ and a joint state $\eta_A$ on $A$ and $\bar{A}$ such that 
for all $x, y \in X(A)$, 
\begin{itemize}
 \item[(a)] $\eta_{A}(x,\bar{x}) = 1/n $
\item[(b)]  $\eta_{A}(x,\bar{y}) = \eta(y,\bar{x})$ 
\end{itemize} 
 where $\bar{x} := \gamma_A(x)$.  

Note that if $E \in \M(A)$, we have $\sum_{x,y \in E \times E} \eta_{A}(x,\bar{y}) = 1$ and $|E| = n$. 
Hence, $\eta_{A}(x,\bar{y}) = 0$ for $x,y \in E$ with $x \not = y$. Thus, $\eta_A$ establishes a 
perfect, uniform correlation between {\em any} test $E \in \M(A)$ and its counterpart, $\bar{E} := \{ \bar{x} | x \in E\}$, in $\M(\bar{A})$. 

The symmetry condition (b) is pretty harmless. If $\eta$ is a joint state on $A$ and $\bar{A}$ satisfying (a), then so is 
$\eta^{t}(x,\bar{y}) := \eta(y,\bar{x})$; thus, $\frac{1}{2}(\eta + \eta^{t})$ satisfies both (a) and (b).  In fact, if $A$ is sharp, 
(b) is automatic: if $\eta$ satisfies (a), then the conditional state $(\eta_{A})_{1|\bar{x}}$ assigns probability one to the outcome $x$. If $A$ is sharp, this implies that $\eta_{1|\bar{x}} = \delta_x$ is uniquely defined, whence, $\eta(x,\bar{y}) = n \delta_{y}(x)$ is also uniquely defined. In other words, for a sharp model $A$ and a given 
isomorphism $\gamma : A \simeq \bar{A}$, there exists 
{\em at most one} joint state $\eta$ satisfying (a) --- whence, in particular, $\eta = \eta^{t}$. 




If $A = A(\H)$ is the quantum-mechanical model associated with an $n$-dimensional Hilbert space $\H$, then we can take $\bar{A} = A(\bar{\H})$ and define $\eta_{A}(x,\bar{y}) = |\langle \Psi, x \otimes \bar{y} \rangle|^2$, where $\Psi$ is the EPR state, as discussed in Section 3. 


So much for conjugates. We generalize the filters associated with pure CP mappings as follows:

{\bf Definition:} A {\em filter} associated with a test $E \in \M(A)$ is a positive linear mapping 
$\Phi : \V(A) \rightarrow \V(A)$ such that for every 
outcome $x \in E$, there is some coefficient 
$t_x \in [0,1]$ with $\Phi(\alpha)(x) = t_x \alpha(x)$ for every state $\alpha \in \Omega(A)$.  

Equivalently, $\Phi$ is a filter iff the dual process $\Phi^{\ast} : \V^{\ast}(A) \rightarrow \V^{\ast}(A)$ satisfies 
$\Phi^{\ast}(\x) = t_x \x$ for each $x \in E$. Just as in the quantum-mechanical case, a filter independently attenuates the ``sensitivity" of the outcomes 
$x \in E$.\footnote{The extreme case is one in which the coefficient $t_{x}$ corresponding to a particular outcome 
is $1$, and the other coefficients are all zero. 
In that case, all outcomes other than $x$ are so to say, blocked by the filter.  Conversely, given such an ``all or nothing" filter $\Phi_x$ for each $x \in E$, we can construct an arbitrary filter with coefficients $t_x$ by setting $\Phi = \sum_{x \in E} t_x \Phi_x$.}

Call a filter $\Phi$ {\em reversible} iff $\Phi$ is an order-automorphism of $\V(A)$; that is,  iff it is probabilistically reversible as a process. 
 Evidently, this requires that all the coefficients $t_x$ be nonzero. 
We'll eventually see that the existence of a  conjugate, plus the {\em preparability} of arbitrary 
nonsingular states by {\em symmetric} reversible filters, will be enough to force $A$ to be a Jordan model. 
Most of the work is done by the easy Lemma 1, below. First, some terminology. 

{\bf Definition:} Suppose $\Delta = \{\delta_{x} | x \in X(A)\}$ is a family of states indexed by outcomes $x \in X(A)$, and such that $\delta_{x}(x) = 1$. Say that a state $\alpha$ is {\em spectral} with respect to $\Delta$ iff there exists a test $E \in \M(A)$ 
such that $\alpha = \sum_{x \in E} \alpha(x) \delta_{x}$. Say that the model $A$ itself 
is spectral with respect to $\Delta$ if every state of $A$ is spectral with respect to $\Delta$. 

If $A$ has a conjugate $\bar{A}$, then the bijection $\bar{(\cdot)} : X(A) \rightarrow X(\bar{A})$ extends to an order-isomorphism $\E(A) \simeq \E(\bar{A})$. It follows that every non-signaling joint probability weight 
$\omega$ on $A$ and $\bar{A}$ defines a bilinear form $a, b \mapsto \omega(a,\bar{b})$ on $\E(A)$. 

{\bf Lemma 1:} {\em Let $A$ have a conjugate $(\bar{A},\eta_{A})$. Suppose $A$ is spectral with respect 
to the states $\delta_{x} := \eta_{1|\bar{x}}$, $x \in X(A)$. Then 
\[\langle a, b \rangle := \eta_{A}(a,\bar{b})\]
defines a self-dualizing inner product on $\E(A)$, with respect to which 
$\V(A)_{+} \simeq \E(A)_{+}$. Moreover, $A$ is sharp, and $\E(A)_{+} = \V^{\ast}(A)_{+}$.}

{\em Proof:} That $\langle \ , \ \rangle$ is symmetric and bilinear follows from $\eta$'s being symmetric and non-signaling. 
We need to show that $\langle \ , \ \rangle$ is positive-definite. Since $\hat{A} \simeq A$, and the latter is spectral, 
so is the former. It follows that $\hat{\eta}$ takes $\E(A)_+$ {\em onto} $\V(A)_+$, and hence, is an order-isomorphism. 
From this, it follows that every $a \in \E(A)_+$ has a ``spectral" decomposition of the form $\sum_{x \in E} t_x x$ for some coefficients $t_x \geq 0$ 
and some test $E \in \M(A)$. In fact, {\em any} $a \in \E(A)$, positive or otherwise, has such a decomposition 
(albeit with possibly negative coefficients).  
For if $a \in \E(A)$ is arbitrary, with $a = a_1 - a_2$ for some $a_1, a_2 \in \E(A)_+$, we can 
find $N  \geq 0$ with $a_2 \leq Nu$. Thus, $b := a + Nu = a_1 + (Nu - a_2) \geq 0$, and so $b := \sum_{x \in E} t_x x$ for some $E \in \A$, 
and hence $a = b - Nu = \sum_{x \in E} t_x x - N(\sum_{x \in E} x) = \sum_{x \in E} (t_x - N)x$. 

Now let $a \in \E(A)$. Decomposing $a = \sum_{x \in E} t_x x$ for 
some test $E$ and some coeffcients $t_x$, we have  
\[
\langle a, a \rangle 
 =  \sum_{x, y \in E \times E} t_x t_y \eta_{A}(x, \bar{y}) 
 =  \frac{1}{n} \sum_{x \in E} {t_{x}}^2 \geq 0.
\]
This is zero only where all coefficients $t_x$ are zero, i.e., only for $a = 0$.  So 
$\langle \ , \ \rangle$ is an inner product, as claimed. 

We need to show that $\langle \ , \ \rangle$ is self-dualizing. Clearly $\langle a, b \rangle = \eta(a, \bar{b}) \geq 0$ for all $a, b \in \E(A)_+$. Suppose $a \in \E(A)$ is such that $\langle a, b \rangle \geq 0$ for all 
$b \in \E(A)_+$. Then $\langle a, \y \rangle \geq 0$ for all $y \in X$. Now, $a = \sum_{x \in E} t_x \x$ for some test $E$; thus, for all $y \in E$ we have 
$\langle a,\y \rangle = t_y \geq 0$, whence, $a \in \E(A)_+$. 

Since $\hat{\eta} : \E(A) \rightarrow \V(\bar{A})$ is an order-isomorphism, for every  
$\alpha \in \V(A)$ there exists a unique $a \in \E(A)$ with $\hat{\eta}(a) = \bar{\alpha}$. 
In particular, 
\[\langle a, x \rangle = \eta(a,\bar{x}) = \bar{\alpha}(\bar{x}) = \alpha(x).\]
It follows that if $b \in \E(A) = \V^{\ast}(A)$, 
\[b(\alpha) = \bar{b}(\bar{\alpha}) = \hat{\eta}(a)(\bar{b}) = \eta(a, \bar{b}) = \langle a, b \rangle.\]
Since every  $a \in \E(A)_+$ has the 
form $a = \hat{\eta}^{-1}(\bar{\alpha})$ for some $\alpha \in \V(A)_+$, 
if $b \in \V^{\ast}(A)_+$, we have $\langle a, b \rangle \geq 0$ for all $a \in \E(A)_+$, whence, 
by the self-duality of the latter cone, $b \in \E(A)_+$.  Thus, $\V^{\ast}(A) = \E(A)_+$.  

Finally, let's see that $A$ is sharp. If $\alpha \in \V(A)$, let $a$ be the unique element of 
$\E(A)_+$ with $\langle a, x \rangle = \alpha(x)$, as discussed above. If $a$ has spectral decomposition $a = \sum_{x \in E} t_x \x$, where $E \in \M(A)$, then 
$\sum_{x \in E} \langle a, x \rangle = \sum_{x \in E} t_x = 1$.  Thus, 
$\|a\|^2 = \sum_{x \in E} t_{x}^{2} \leq 1$, whence, $\|a\| \leq 1$. Now suppose 
$\alpha(x) = 1$ for some $x \in X(A)$: then $1 = \langle a, x \rangle \leq \|a\| \|x\|$; 
as $\|x\| = 1$, we have $\|a\| = 1$. But now $\langle a, 
\hat{x} \rangle = \|a\| \|\x\|$, whence, 
$a = \x$. Hence, there is only one weight $\alpha$ with $\alpha(x) = 1$, namely, 
$\alpha = \langle x, \, \cdot \, \rangle$, so $A$ is sharp. 
$\Box$  




If $A$ is sharp, then we say that $A$ is {\em spectral}  iff it is spectral with respect to the pure states $\delta_x$ defined by $\delta_x(x) = 1$.  If $A$ is sharp and has a conjugate $\bar{A}$, then, as noted earlier, the state $ \eta_{1|\bar{x}}$ is exactly $\delta_x$, so the spectrality assumption in Lemma 1 is fulfilled if we simply say that $A$ is spectral. 
Hence, {\em a sharp, spectral model with a conjugate is self-dual.}

For the simplest systems, this is already enough to secure the desired representation in terms of a euclidean Jordan algebra. 

{\bf Definition:} Call $A$ a {\em bit}  
iff it has rank $2$ (that is, all tests have two outcomes), and if every state $\alpha \in \Omega(A)$ can be expressed 
as a mixture of two sharply distinguishable states; that is, $\alpha = t \delta_x + (1 - t)\delta_y$ for some $t \in [0,1]$ and states $\delta_x$ and $\delta_y$ 
with $\delta_x(x) = 1$ and $\delta_y(y) = 1$ for some test $\{x,y\}$. 


{\bf Corollary 1:} {\em If $A$ is a sharp bit, then $\Omega(A)$ is a ball of some finite dimension $d$.} 

The proof is given in Appendix C. 
If $d$ is $2, 3$ or $5$, we have a real, complex or quaternionic bit. For $d = 4$ or $d \geq 6$, we have a non-quantum spin factor.


For systems of higher rank (higher ``information capacity"), we need to assume a bit more. Suppose $A$ satisfies the hypotheses of Lemma 1. Appealing to the Koecher-Vinberg Theorem,  we 
see that if $\V(A)$ and, 
hence, $\V^{\ast}(A)$ are also homogeneous, then $\V^{\ast}(A)$ carries a canonical Jordan structure. In fact, we can say something a little stronger. 

{\bf Theorem 1:} {\em Let $A$ be spectral with respect to a conjugate system $\bar{A}$.  
If $\V(A)$ is homogeneous, then there exists a canonical Jordan product on $\E(A)$ with respect to which $u_A$ is the Jordan unit. Moreover with respect to this product $X(A)$ is exactly the set of primitive idempotents, and $\M(A)$ is exactly the set of Jordan frames.} 


The first part is almost immediate from the Koecher-Vinberg Theorem, together with Lemma 1. The KV Theorem gives us an isomorphism between the ordered vector spaces $\V(A)$ and 
$\E(A)$, so if one is homogeneous, so is the other. Since $\E(A)$ is also self-dual by Lemma 1, 
the KV theorem yields the requisite unique euclidean Jordan structure having $u$ as the Jordan unit. 
One can then show without much trouble that every outcome $x \in X(A)$ is a primitive 
idempotent of $\E(A)$ with respect to this Jordan structure, and that every test is a Jordan frame.  
The remaining claims 
---  that every minimal idempotent belongs to $X(A)$ and every Jordan frame, to $\M(A)$ 
--- take a little bit more work. I won't reproduce the proof here; the details 
(which are not difficult, but depend on some facts concerning euclidean Jordan algebras) 
can be found in \cite{Wilce12}.

The homogeneity of $\V(A)$ can be understood as a {\em preparability assumption:} it is equivalent 
to saying that every state in the interior of $\Omega(A)$ can be obtained, {\em up to normalization}, from the 
maximally mixed state by a reversible process. That is, if $\alpha \in \Omega(A)$, there is some 
such process $\phi$ such that $\phi(\rho) = p \alpha$ where $0 < p \leq 1$. One can think of the coefficient 
$p$ as the probability that the process $\phi$ will yield a nonzero result (more dramatically: will not 
destroy the system). Thus, if we prepare an ensemble of identical copies of the system in the maximally mixed state $\rho$ and subject them all to the process $\phi$, the fraction that survive will be about $p$, and these 
will all be in state $\alpha$. 

In fact, if the hypotheses of Lemma 1 hold, the homogeneity of $\E(A)$ follows directly from the mere existence of reversible filters with arbitrary non-zero coefficients. To see this, suppose $a \in \E(A)_+$ has a spectral decomposition $\sum_{x \in E} t_x \x$ for some $E \in \M(A)$, with $t_x > 0$ for all $x$ when $a$ belongs to the 
interior of $\E(A)_+$. Now if we can find a reversible filter for $E$ with $\Phi(x) = t_x \x$ for all $x \in E$, then applying this to the order-unit $u = \sum_{x \in E} \x$ yields $a$. Thus, $\V^{\ast}(A)$ is homogeneous. 

{\bf Two paths to spectrality}   Some axiomatic treatments of quantum theory have taken one or another form of spectrality as an axiom \cite{Gunson, BMU}. If one is content to do this, then Lemma 1 above provides a very direct route to the Jordan structure of quantum theory. However, spectrality can actually be derived from assumptions that, on their face, seem a good deal weaker, or anyway more transparent.\footnote{A different path to spectrality 
is charted in a recent paper \cite{CS} by G. Chiribella and C. M. Scandolo.}


I will call a joint state on models $A$ and $B$ {\em correlating} iff it sets up a perfect correlation 
between {\em some} pair of tests $E \in \M(A)$ and $F \in \M(B)$. More exactly:

{\bf Definition:} A joint state $\omega$ on probabilistic models $A$ and $B$ {\em correlates} 
a test $E \in \M(A)$ with a test $F \in \M(B)$ iff there exist subsets $E_{0} \subseteq E$ and $F_{0} \subseteq F$, 
and a bijection $f : E_{0} \rightarrow F_{0}$ such that $\omega(x,y) = 0$ for $(x,y) \in E \times F$ 
unless $y = f(x)$. In this case, say that $\omega$ correlates $E$ with $F$ {\em along $f$}.  A joint state on $A$ and $B$ is {\em correlating} iff it correlates some pair 
of tests $E \in \M(A), F \in \M(B)$. 

Note that $\omega$ correlates $E$ with $F$ along $f$ iff 
$\omega(x,f(x)) = \omega_1(x) = \omega_2(f(x))$, which, in turn, 
is equivalent to saying that 
$\omega_{2|x}(f(x)) = 1$ for $\omega_1(x) \not = 0$.

{\bf Lemma 2:} {\em Suppose $A$ is sharp, and that every state $\alpha$ of $A$ arises as the marginal 
of a correlating joint state between $A$ and some model $B$. Then $A$ is spectral.}

{\em Proof:} Suppose $\alpha = \omega_1$, where $\omega$ is a joint state correlating 
a test $E \in \M(A)$ with a test $F \in \M(B)$, say along an bijection $f : E_{0} \rightarrow F_{0}$, 
$E_o \subseteq E$ and $F_{0} \subseteq F$. Then for any $x \in E$ with $\alpha(x) \not = 0$, 
$\omega_{1|f(x)}(x) = 1$, whence, as $A$ is sharp, $\omega_{1|f(x)} = \delta_x$, the unique 
state making $x$ certain. It follows from the Law of Total Probability that 
$\alpha = \sum_{x \in E} \alpha(x) \delta_x$. $\Box$ 

In principle, the model $B$ can vary with the state $\alpha$. Lemma 2 suggests the following language:


{\bf Definition:} A model $A$ satisfies the {\em correlation condition} iff every state $\alpha \in \Omega(A)$ 
is the marginal of some correlating joint state of $A$ and some model $B$.

This has something of the same flavor as the {\em purification postulate} of \cite{CDP}, which requires that 
all states of a given system arise as marginals of a {\em pure} state on a larger, composite system, unique 
up to symmetries on the purifying system. However, note that we do not require the correlating joint state 
to be either pure (which, in classical probability theory, it will not be), or unique.  

 
If $A$ is sharp and satisfies the correlation condition, then every state of $A$ is spectral. If, in addition, $A$ has 
a conjugate, then for every $x \in X(A)$, we have 
$\eta_{1|\bar{x}} = \delta_{x}$. In this case, $A$'s states are spectral with respect to the 
family of states $\eta_{1|\bar{x}}$, and the hypotheses of Lemma 1 are satisfied. 
So we have 


\tempout{{\em Proof:} (a) Suppose that a state $\alpha$ of $A$ arises as the marginal of a joint state $\omega$ that 
correlates a test $E \in \M(A)$ with its counterpart $\bar{E} := \{ \bar{x} | x \in E\} \in \M(\bar{A})$. This means that $\omega(x,\bar{y}) = 0$ for $x, y \in E$ with $x \not = y$. It follows that 
$\omega_{1|\bar{x}}(x) = 1$.  Since $A$ is sharp, $\omega_{1|\bar{x}} = \eta_{1|\bar{x}} := \delta_x$. The Law of Total Probability (1) gives us $\alpha = \omega_1 = \sum_{x \in E} \omega_{2}(\bar{x}) \delta_{x}$, which is 
just the spectral decomposition required. }

Here is another, superficially quite different, way of arriving at spectrality. 
Call a transformation $\Phi$ {\em symmetric} with respect to $\eta_A$ iff, 
for all $x, y \in X(A)$, 
\[\eta_{A}(\Phi^{\ast} x, \bar{y} ) = \eta_{A}(x, \bar{\Phi}^{\ast} y).\]
 
{\bf Lemma 3:} {\em Let $A$ have a conjugate, $\bar{A}$, and suppose every state of $A$ is preparable by a 
symmetric filter. Then $A$ is spectral.}  
 
{\em Proof:} Let $\alpha = \Phi(\rho)$ where $\Phi$ is a filter on a test $E \in \M(A)$, say $\Phi(x) = t_x x$ for all $x \in E$.
Then 
\[\ \alpha = \Phi(\hat{\eta}^{\ast}(\bar{u})) = \eta(\Phi^{\ast}( \cdot ), \bar{u}) = \eta( \ \cdot \ , \bar{\Phi}^{\ast}(\bar{u})) = 
\sum_{x \in E} \eta( \ \cdot \ , t_x \bar{x}) = \sum_{x \in E} t_{x} \tfrac{1}{n} \delta_{x}. \ \ \Box\]

Thus, the hypotheses of either Corollary 2 or Lemma 3 will supply the needed spectral assumption that makes Lemma 1 
work. (In fact, it is not hard to see that these hypotheses are actually equivalent --- an exercise I leave for 
the reader.) 

To obtain a {\red Jordan model}, we still need homogeneity. This is obviously implied by the preparability condition in Lemma 3, provided the preparing filters $\Phi$ can be taken to be reversible whenever the 
state to be prepared is non-singular. On the other hand, as noted above, in the presence of spectrality, it's enough to {\em have} arbitrary reversible filters, as these allow one to prepare the spectral decompositions of arbitrary non-singular states. Thus, conditions (a) and (b), below, both imply that $A$ is a Jordan model. Conversely, one can show that any Jordan model satisfies 
both (a) and (b), closing the loop \cite{Wilce12}: 


{\bf Theorem 2:} {\em The following are equivalent: 
\begin{mlist} 
\item[(a)] $A$ has a conjugate, and every non-singular state can be prepared by a reversible symmetric filter;
\item[(b)] $A$ is sharp, has a conjugate, satisfies the correlation condition, and has arbitary reversible filters; 
\item[(c)] $A$ is a Jordan model.
\end{mlist}}

\section{Measurement, Memory and Correlation}

Of the spectrality-underwriting conditions given in Lemmas 2 and 3, the one that seems less transparent (to me, anyway) is  that every state arise as the marginal of a correlating biparite state. 
While surely less {\em ad hoc} than  spectrality, this still calls for further explanation. 
Suppose we hope to implement a measurement of a test $E \in \M(A)$ {\em dynamically}. 
This would involve bringing up an ancilla system $B$ --- also uniform, suppose; and which we can suppose (by suitable coarse-graining, if necessary) to have tests of the same cardinality as $A$'s --- in some ``ready" state $\beta_o$. 
We would then subject the combined system $AB$ to some physical process, at the end of which $AB$ is in some 
final joint state $\omega$, and $B$ is (somehow!) in one 
of a set of {\em record states}, $\beta_x$, each corresponding to an outcome $x \in X(A)$.\footnote{
This way of putting things takes us close to the usual formulation of the quantum-mechanical ``measurement problem", 
which I certainly don't propose to discuss here. The point is only that, if {\em any} dynamical process, describable 
within the theory, can account for measurement results, it should be consistent with this description.}
We would like to 
insist that 
\begin{mlist} 
\item[(a)] The states $\beta_x$ are {\em distinguishable}, or {\em readable}, by some test $F \in \M(B)$. This 
means that for each $x \in E$, there is a unique $y \in F$ such that $\beta_{x}(y) = 1$. Note that this 
sets up an injection $f : E \rightarrow F$. 
\item[(b)] The record states must be {\em accurate}, in the sense that if we {\em were} to measure 
$E$ on $A$, and secure $x \in E$, the record state $\beta_x$ should coincide with the conditional state 
$\omega_{2|x}$. (If this is not the case, then a measurement of $A$ cannot correctly calibrate $B$ as a 
measuring device for $E$.) 
\end{mlist} 
It follows from (a) and (b) that, for $x \in E$ and $y \not = f(x) \in F$, 
\[\omega(x,y) = \omega_1(x) \omega_{2|x}(y) = \omega_1(x) \beta_{x}(y) = 0.\]
In other words, $\omega$ must correlate $E$ with $F$, along the bijection $f$. If the measurement process 
leaves $\alpha$ undisturbed, in the sense that $\omega_1 = \alpha$, then $\alpha$ dilates to a correlating 
state. This suggests the following {\em non-disturbance principle:} every state can be measured, by some test 
$E \in \M(A)$, without disturbance. 
Lemma 2 then tells us that if $A$ is sharp and satifies the non-disturbance principle, every state of $A$ is spectral.  

Here is a slightly different, but possibly more compelling, version of this story. Suppose we can 
make a test $E$ on $A$ directly (setting aside, that is, any issue of whether or not this can 
be achieved through some dynamical process): this will result in an outcome $x$ {\em occuring}. To {\em do} 
anything with this, we need to {\em record} its having occurred. This means we need a storage medium, 
$B$, and a family of states $\beta_x$, one for each $x \in E$, such that {\em if}, on performing the test 
$E$, we obtain $x$, then $B$ {\em will} be in state $\beta_x$. Moreover, these record states need to be {\em readable} at 
a later time, i.e.,  distinguishable by a later measurement on $B$.  
To arrange this, we need $A$ and $B$ to be in a joint state, associated with a joint probability weight $\omega$, such that $\omega_1 = \alpha$ (because we 
want to have prepared $A$ in the state $\alpha$) and $\beta_x = \omega_{2|x}$ for every $x \in E$. We then measure $E$ on $A$; 
upon our obtaining outcome $x \in E$, $B$ is in the state $\beta_{x}$. 
Since the ensemble of states $\beta_x$ is readable (by some $F \in \M(B)$ with $|F| \geq |E|$), we have correlation, and $\alpha$ must also be spectral. 

Of course, these desiderata can not always be satisfied. What {\em is} true, in QM, is that for every choice of 
state $\alpha$, there will exist {\em some} test that is recordable that state, in the foregoing sense.  If we  
promote this to general principle, we again see that every state is the marginal of a correlating state, and hence   spectral, if $A$ is sharp.

\tempout{
We can understand the conjugate system $\bar{A}$ in these terms: if $A$ and $\bar{A}$ are in a joint state associated 
with the correlator $\eta_A$, then {\em any} measurement made on $A$ in the maximally mixed state 
$\bar{A}$. This means, 
in practice, that after Alice (say) performs test $E \in \M(A)$, the conjugate system is left in a record state 
such that, {\em if} Bob performs the same test (or any test containing the outcome Alice obtained), he will 
obtain the ``correct" result, i.e., the same result Alice got. }




\section{Composites and Categories} 

Thus far, we've been referring to the correlator $\eta_A$ as a {\em joint state}, but dodging the question: 
{\em state of what?} 
Mathematically, 
nothing much hangs on this question: it 's sufficient to regard $\eta_A$ as a bipartite probability assignment 
on $A$ and $\bar{A}$. But it would surely be more satisfactory to be able to treat it as an actual physical 
state of some composite system $A \bar{A}$. How should this be chosen? 

One possibility is to take $A \bar{A}$ to be the {\em maximal} tensor product of the models $A$ and $\bar{A}$ 
 \cite{BarnumWilceFoils}. By definition, this has for its states all nonsignaling probability assignments with conditional 
states belonging to $A$ and $\bar{A}$. 
However, we might want composite systems, in particular $A \bar{A}$, to satisfy the same conditions we are imposing on $A$ and $\bar{A}$, i.e., to be a Jordan model. If so, we need to work somewhat harder: the maximal tensor product will 
be self-dual only if $A$ is classical.

In order to be more precise about all this, the first step is to decide what ought to count as a composite of two probabilistic models.  If we mean to capture the idea of two physical 
systems that can be acted upon separately, but which cannot influence one another in any observable way (e.g., two spacelike-separated systems), the following seems to me to capture the minimal requirements:

{\bf Definition:} A {\em non-signaling composite} of models $A$ and $B$ is a model $AB$, together with a mapping $\pi : X(A) \times X(B) \rightarrow \V^{\ast}(AB)_{+}$ such that 
\[\sum_{x \in E, y \in F} \pi(x,y) = u_{AB}\]
and, for $\omega \in \Omega(AB)$, $\omega \circ \pi$ is a joint state on $A$ and $B$, as defined in Section 2.

So the question becomes: can one construct, for Jordan models $A$ and $B$, a nonsignaling composite $AB$ that is also a Jordan model? At present, and in this generality, this question seems to be open, but 
some progress is made in \cite{CCEJA}. I will have more to say about this below. 
As shown in \cite{CCEJA}, neither $A$ nor $B$ contains the exceptional Jordan algebra as a summand, such a composite can indeed be constructed, and in multiple ways.  Under a considerably more 
restrictive definition of ``Jordan composite", it is also shown in \cite{CCEJA} that no Jordan composite $AB$ can exist if either factor has an exceptional summand. 



{\bf Categories of Self-Dual Probabilistic Models} \ 
It's natural to interpret a physical theory as a {\em category}, in which objects represent physical systems and morphisms represent physical processes having these 
systems (or there states) as inputs and outputs.  In order to discuss composite systems, this should be a {\em symmetric monoidal} category. That is, for every pair of objects 
$A, B$, there should be an object $A \otimes B$, and for every pair of morphisms $f : A \rightarrow A'$ and $g : B \rightarrow B'$, there should be a morphism 
$f \otimes g : A \otimes B \rightarrow A' \otimes B'$, representing the two processes $f$ and $g$ occuring ``in parallel". One requires that 
$\otimes$ be associative and commutative, and have a unit object $I$, in the sense that there exist canonical isomorphisms 
$\alpha_{A,B;C} : A \otimes (B \otimes C) \simeq (A \otimes B) \otimes C$,  
$\sigma_{A,B} : A \otimes B \simeq B \otimes A$, $\lambda_{A} : I \otimes A \simeq A$ 
and 
 $\rho_A : A \otimes I \rightarrow A$/ 
These must atisfy various ``naturality conditions", guaranteeing that they interact correctly; see \cite{MacLane} for details. One also requires that $\otimes$ be bifunctorial, 
meaning that $\id_{A} \otimes \id_{B} = \id_{A \otimes B}$, and  
if $f : A \rightarrow A'$, $f' : A' \rightarrow A''$, $g : B \rightarrow B'$ and $g' : B' \rightarrow B''$, then 
\[(f' \otimes g') \circ (f \otimes g) = (f' \circ f) \otimes (g' \circ g).\]

Following this lead, by a {\em probabilistic theory}, I mean a category  of probabilistic models and processes --- that is, objects of $\Cat$ are models, and a morphism $A \rightarrow B$, where 
$A, B \in \Cat$, is a process $\V(A) \rightarrow \V(B)$ 
A {\em monoidal} probabilistic theory is such a category, $\Cat$, carrying a symmetric monoidal structure $A, B \mapsto AB$, where $AB$ is a non-signaling composite in the sense of 
the Definition above. I also assume that the monoidal unit, $I$, is the trivial model $1$ with $\V(1) = \R$, 
and that, for all $A \in \Cat$, 
\vspace{.03in}
\begin{mlist}
\item[(b)] $\alpha \in \Omega(A)$ iff the mapping $\alpha : \R \rightarrow \V(A)$ given by $\alpha(1) = \alpha$ belongs to $\Cat(I,A)$; 
\item[(b)] The evaluation functional $\x$ belongs to $\Cat(A,I)$ for all outcomes $x \in X(A)$.
\end{mlist}
\vspace*{-.12in}
Call $\Cat$ {\em locally tomographic} iff $AB$ is a locally-tomographic composite for all $A, B \in \Cat$; 

Much of the qualitative content of (finite-dimensional) quantum information theory can be formulated 
in purely categorical terms \cite{Abramsky-Coecke, Baez, Selinger}. In particular, in the work of Abramsky and Coecke \cite{Abramsky-Coecke}, it is shown that a range of 
quantum phenomena, notably gate teleportation, are available in any {\em dagger-compact} category.  
For a review of this notion, as well as a proof of the following result, see Appendix D:


{\bf Theorem 3:} {\em Let $\Cat$ be a locally tomographic monoidal probabilistic theory, in which every object 
$A \in \Cat$ is 
sharp, spectral, and has a conjugate $\bar{A} \in \Cat$, 
with $\eta_A \in \Omega(A \bar{A})$. Assume also that, for all $A, B \in \Cat$,
\vspace{.02in} 
\begin{mlist} 
\item[(i)] $\bar{\bar{A}} = A$, with $\eta_{\bar{A}}(\bar{a},b) = \eta_{A}(a,\bar{b})$; 
\item[(ii)]  If $\phi \in \Cat(A,B)$, then $\bar{\phi} \in \Cat(\bar{A},\bar{B})$.
\end{mlist}
\vspace{-.1in}
Then $\Cat$ has a canonical dagger-compact structure, in which $\bar{A}$ is the 
dual of $A$ with $\eta_{A} : \R \rightarrow \V(A\bar{A})$ as the co-unit.}


{\bf Jordan Composites}  The local tomography assumption in Theorem 3 is a strong constraint. As is well known, the standard composite of two real quantum systems 
is not locally tomographic, yet the category of finite-dimensional real mixed-state quantum systems is certainly dagger-compact, and satisfies the other assumptions 
of Theorem 3, so local tomography is definitely not a necessary condition for dagger-compactness.  

This raises some questions. One is whether local tomography can simply be dropped in the statement of Theorem 3.  At any rate, at present I don't know of any non-dagger-compact monoidal probabilistic theory satisfying the other assumptions. 

Another question is whether there exist examples {\em other} than real QM of non-locally tomographic, but still dagger-compact, monoidal probabilistic theories satisfying the assumptions of Theorem 1.  The answer to this is {\em yes}.  
Without going into detail, the main result of \cite{CCEJA} is that one can construct a dagger-compact category in which the objects are hermitian parts of finite-dimensional real, complex and quaternionic matrix algebras --- that is, the euclidean 
Jordan algebras corresponding to finite-dimensional real, complex or quaternionic quantum-mechanical systems --- and morphisms are certain completely positive mappings between 
enveloping complex $\ast$-algebras for these Jordan algebras. The monoidal structure gives {\em almost} the expected results: the composite of two real quantum systems 
is the real system corresponding to the usual (real) quantum-mechanical composite of the two components (and, in particular, is not locally tomographic). The composite of two quaternionic systems is a real system (see \cite{Baez} for a 
account of why this is just what one wants). The composite of a real and a complex, or a quaternionic and a complex, system is again complex. The one surprise 
is that the composite of two standard {\em complex} quantum systems, in this category, is {\em not} the usual thing, but rather, comes with an extra superselection rule. 
This functions to make time-reversal a legitimate physical operation on complex systems, as it is for real and quaternionic systems. This is part of the price one pays 
 for the dagger-compactness of this category. 

\section{Conclusion} 

As promised, we have here an easy derivation of something {\em close} to orthodox, finite-dimensional QM, from operationally or probabilistically transparent assumptions. As discussed earlier, this approach offers --- in addition to its relative simplicity --- greater lattitude than the 
locally tomographic axiomatic reconstructions of \cite{Hardy, Dakic-Brukner, Masanes-Mueller, CDP},  
putting us in the slightly less constrained realm of formally real Jordan algebras. This  allows for real and quaternionic 
quantum systems, superselection rules, and even theories, such as the ones discussed in section 6, in which real, complex and quaternionic quantum systems coexist and interact.

There remains some mystery as to the proper interpretation of the conjugate system $\bar{A}$. Operationally, 
the situation is clear enough: if we understand $A$ as controlled by Alice and $\bar{A}$, by Bob, then 
if Alice and Bob share the state $\eta_A$, then they will always obtain the same result, as long as they perform the 
same test. But what does it mean {\em physically} that this should be possible (in a situation in which 
Alice and Bob are still able to chose their tests indpendently)?               
In fact, there is little consensus (that I can find, anyway) among physicists as to the proper interpretation 
of the conjugate of the Hilbert space representing a given quantum-mechanical system. One popular 
idea is that the conjugate is a {\em time-reversed} version of the given system --- but why, then, should 
we expect to find a state that perfectly correlates the two? At any rate, finding a clear physical intepretation 
of conjugate systems, even --- or especially! --- in orthodox quantum mechanics, seems to me an important 
problem. 



I'd  like to close with another problem, this one of mainly mathematical interest. The hypotheses 
of Theorem 1 yield a good deal more structure than just an homogeneous, self-dual cone. In particular, 
we have a distinguished set $\M(A)$ of orthonormal observables in $\V^{\ast}(A)$, with respect to which every 
effect has a spectral decomposition. Moreover, with a bit of work one can show that this decomposition is 
essentially unique. More exactly, if $a = \sum_i t_i p_i$ where the coefficients $t_i$ are all 
distinct and the effects $p_1,...,p_k$ are associated with a coarse-graining of a test $E \in \M(A)$, 
then both the coefficients and the effects are uniquely determined. The details are in Appendix B. Using this, we have a functional 
calculus on $\V^{\ast}(A)$, i.e., for any real-valued function $f$ of a real variable, 
and any effect $a$ with spectral decomposition $\sum_i t_i p_i$ as above, 
we can define $f(a) = \sum_i f(t_i) p_i$. 
This gives us a unique candidate for the Jordan product of effects $a$ and $b$, namely, 
\[a \dot b = \tfrac{1}{2}((a + b)^2 - a^2 - b^2)).\]
We know from Theorem 1 --- and thus, ultimately, from the KV theorem --- that this is bilinear.
The challenge is to show this {\em without} appealing to the KV theorem.  (The fact that 
the state spaces of ``bits" are always balls, as shown in Appendix C, is perhaps relevant here.)


\tempout{
A second problem is to extend the results discussed here to infinite-dimensional systems. The KV theorem 
is an inherently finite-dimensional result, and our definition of a conjugate system, which depends on the existence of a 
uniformly mixed state, requires tests of a fixed {\em finite} size. ...

While the definition of a conjugate system given above depends on the existence of a maximally mixed state, and is thus applicable only to models 
having finite tests of a common size, the definition of an equivariant conjugate makes sense for infinite-dimensional systems. Thus, if we could bypass the KV theorem, it might be possible 
also to jettison the assumption that $\V^{\ast}(A)$ is finite-dimensional. 
}

{\bf Acknowledgements} This paper is partly based on talks given in workshops and seminars in Amsterdam, 
Oxford in 2014 and 2015, and was largely written while the author was a guest of the Quantum Group at the Oxford Computing Laboratory, supported by a grant (FQXi-RFP3-1348) from the FQXi foundation. I would like to thank Drs. Sonja Smets (in Amsterdam) and Bob Coecke (in Oxford) for their hospitality on these occasions. I also wish to thank Carlo Maria Scandolo for his careful reading of, and useful comments on, two earlier drafts of this paper. 

\appendix
\section{Models with symmetry} 

Recall that a probabilistic model $A$ is {\em sharp} iff, for every measurement outcome 
$x \in X(A)$, there exists a {\em unique} state $\delta_x \in \Omega(A)$ with $\delta_x(x) = 1$. While this 
is clearly a very strong condition, it is not an unreasonable one.  In fact, given the test space 
$\M(A)$, we can often 
{\em choose} the state space $\Omega(A)$ in such a way as to guarantee that $A$ is sharp.  In particular, 
this is the case when $\M(A)$ enjoys enough symmetry. 

{\bf Definition:} Let $G$ be a group. A {\em $G$-test space} is a test space $(X,\M)$ where $X$ is a $G$-space --- that is, where $X$ comes equipped with a preferred $G$-action $G \times X \rightarrow X$, $(g,x) \mapsto gx$ --- such that 
$gE \in \M$ for all $E \in \M$.  A {\em $G$-model} is a probabilistic model 
$A$ such that (i) $\M(A)$ is a $G$-test space, and (ii) $\Omega(A)$ is invariant under the action of $G$ on probability 
weights given by $\alpha \mapsto g\alpha := \alpha \circ g^{-1}$ for $g \in G$.



{\bf Lemma A1:} {\em Let $A$ be a finite-dimensional $G$-model and suppose $G$ acts transitively on the outcome space $X(A)$. 
Suppose also that $A$ is unital, i.e., for every $x \in X(A)$, there exists at least one state $\alpha$ with $\alpha(x) = 1$. Then there exists a $G$-invariant convex subset $\Delta \subseteq \Omega(A)$ such that $A' = (\M(A),\Delta)$ is a sharp $G$-model.}

{\em Proof:} For each $x \in X(A)$, let $F_x$ denote the face of $\Omega(A)$ consisting of states $\alpha$ with $\alpha(x) = 1$. Let $\beta_x$ be the barycenter of $F_x$. It is easy to check that $F_{gx} = gF_{x}$ for every $g \in G$. Thus, $g\beta_x = \beta_{gx}$, i.e., the set of barycenters $\beta_x$ is an orbit. Let $\Delta$ be the convex hull of these barycenters. Then $\Delta$ is invariant under $G$. If $\alpha \in \Delta$ with $\alpha(x) = 1$, then $\alpha \in F_{x} \cap \Delta = \{\beta_x\}$, so $(\M(A),\Delta)$ is sharp. $\Box$.

\section{Uniqueness of Spectral Decompositions} 

Let $A$ be a model satisfying the conditions of Lemma 1. In particular, then, every $a \in \E(A) = \V^{\ast}(A)$ has a spectral representation $a = \sum_{x \in E} t_x \x$ for some test $E \in \M(A)$. In general, this expansion is highly non-unique. For instance, the unit $u_A$ can be expanded as $\sum_{x \in E} \x$ for any test $E \in \M(A)$. The aim 
in this Appendix is to obtain a form of spectral expansion for effects that {\em is} unique. 

Call a subset of a test an {\em event}.  That is, $D \subseteq X(A)$ is an event iff there exists 
a test $E \in \M(A)$ with $D \subseteq E$. Any event gives rise to an effect 
\[p(D) := \sum_{x \in D} \x.\] 
A test is a maximal event, and for any test  $E \in \M(A)$, $p(E) = u$.

Let's say that an effect $a \in \V^{\ast}(A)$ is {\em sharp} iff it has the form $p(D)$ for some 
event $D$.  

{\bf Definition:} A set of sharp effects $p_1,...,p_n \in \V^{\ast}(A)$ is {\em jointly orthogonal} with respect to $\M(A)$ iff there exists a test $E \in \M(A)$ and pairwise disjoint events $D_1,...,D_n \subseteq E$ with $p_i = p(D_i)$ for $i = 1,...,n$. 

Given an arbitrary element $a \in \V^{\ast}(A)$ with spectral decomposition 
$a = \sum_{x \in E} t_x \x$, we can isolate distinct values $t_o > t_1 > ... > t_k$ of the coefficients 
$t_x$. Letting $E_i = \{ x \in E | t_x = t_i\}$ and setting $p_i = p(E_i) = \sum_{x \in E_i} \x$, we have 
$a = \sum_i t_i p_i$, with $p_1,...,p_n$ jointly orthogonal. Suppose there is another such decomposition, say $a = \sum_j s_j q_j$, with $q_j = p(F_j) = \sum_{y \in F_j} \y$, where $F_1,...,F_l \subseteq F \in \M(A)$ are pairwise disjoint, 
and again with the coefficients in descending order, say $s_0 > s_1 > \cdots > s_l$.  

{\bf Lemma B1:} {\em In the situation described above, $t_0 = s_0$ and $p_0 = q_0$.} 


{\em Proof:} Normalize the inner product on $\E(A)$ so that $\|x\| = 1$ for all outcomes $x$. 
Then 
for any choice of outcome $x_o \in E$, set $\alpha = |x_o\rangle$, 
i.e., $\alpha(\x) = \langle \x, \x_o \rangle$ for all $x \in X(A)$, 
we have $\alpha(p_0) = 1$ and $\alpha(p_i) = 0$ for $i > 0$. Thus, 
\[t_o = \alpha(a) = \sum_{j} s_j \alpha(q_j).\]
Since the coefficients $\alpha(q_j)$ are sub-convex, the right-hand side 
is no larger than the largest of the values $s_j$, namely,  
$s_o$. Thus, $t_0 \leq s_0$. A similar argument shows that $s_0 \leq t_0$. Thus, $s_0 = t_0$. 

Now again let $x \in E$: then 
\[\langle \x, p_0 \rangle = \sum_{y \in E_0} \langle \x, \y \rangle = \langle x, x \rangle = 1,\]
whence, 
$\langle \x, a \rangle = t_o$.  But we then have (using the fact that $s_0 = t_0$) 
\[t_0 = \langle \x, a \rangle = \left \langle \x \ , \ t_0 q_0 + \sum_{j=1}^{l} s_j q_j \right \rangle = t_o \langle \x, q_0 \rangle + \sum_{j=1}^{l} s_j \langle \x,  q_j \rangle.\]
Since $\sum_{j=0}^{l} \langle \x, q_j \rangle \leq 1$, the sum in the last expression above is a sub-convex combination of the distinct values 
$s_o > \cdots  > s_l$. This can equal $t_0 = s_0$, the maximum of these values, only if $\langle \x, q_0 \rangle = 1$ and $\langle \x, q_j \rangle = 0$ for the remaining 
$q_j$. It follows that $\langle p_0, q_0 \rangle = \sum_{x \in E_0} \langle \x, q_0 \rangle = |E_0| = \|p_0\|^2$. The same argument, with $p$'s and $q$'s interchanged, 
shows that $\langle p_0, q_0 \rangle = \|q_0\|^2$. Hence, $\|p_0\| = \|q_0\|$, and $\langle p_0, q_0 \rangle = \|p_0\|^2 = \|p_0\|\|q_0\|$, whence, $p_0 = q_0$ $\Box$

\tempout{
Hence, 
\[\|p_o\|^2 = \langle p_o, p_o \rangle =\sum_{x \in E_o} \langle \x, p_o \rangle = |E_{o}|.\]
As the effects $p_i$ are pairwise orthogonal with respect to the inner product on $\V^{\ast}(A)$. 
$\langle \x, a \rangle = t_o$. We now have 
Now notice that as the effects $p_i$ are orthogonal to one another with respect to the inner product 
on $\V^{\ast}(A)$, we have 
\[\|p_o\|^2 = \langle p_o, p_o \rangle = \langle p_o, a \rangle = t_o\]
and similarly 
\[\|q_o\|^2 = \langle q_o, q_o \rangle = \langle q_o, a \rangle = s_o.\]
Since $t_o = s_o$, $\|p_o\|^2 = \|q_o\|^2$, whence, $\|p_o\| = \|q_o\|$. 

Now let $x \in E_o$ and note that $\langle \x | a \rangle = t_o = s_o$, so that 
with $\alpha = \langle x |$, 
$s_o = \langle x | a \rangle = \sum_{j} \alpha(q_j)s_j$. The convexity of the 
coefficients $\alpha(q_j)$ again implies 
$\alpha(q_o) = 1$ and $\alpha(q_j) = 0$ for $j > 0$. 
It follows, then, that 
$\alpha(q') = 0$ for $q' = u - q_o$. Since this holds for all $x \in E_o$, we see 
that $\sum_{x \in E_o} \x = p_o$ also satisfies $\langle p_o, q_{o}' \rangle = 0$. Hence, 
\[\|p_o\|^2 = \langle p_o, u \rangle = \langle p_o, q_o \rangle \leq \|p_o\|\|q_o\|.\]
Since $\|q_{o}\| = \|p_o\|$, it follows that the last inequality is an equality, whence,  
$p_o = q_o$. $\Box$ }

{\bf Proposition B2:} {\em Every $a \in \V^{\ast}(A)$ has a unique expansion of the form 
$a = \sum_{i=0}^{k} t_i p_i$ 
where $t_0 > t_1 > ... > t_k$ are non-zero coefficients and $p_1,...,p_n$ are jointly orthogonal sharp 
effects.} 

{\em Proof:} Suppose $a = \sum_{i=1}^{k} t_i p_i$, as above, and also $a = \sum_{j=1}^{l} s_j q_j$, 
$s_0 > \cdots > s_l > 0$,  and $q_j$ pairwise orthogonal sharp effects. We shall 
show that $k = l$, and that $t_i = s_i$ and $p_i = q_i$ for each $i = 1,...,k$. 
Lemma B1  tells us that $t_0 = s_0$ and $p_0 = s_0$. Hence, 
\[\sum_{i = 1}^{k} t_i p_i = a - t_o p_o = a - s_0 q_0 = \sum_{j=1}^{l} s_j q_j.\] 
Applying Lemma B1 recursively, we find that $t_i = s_i$ and $p_i = q_i$ 
for $i = 1,...,\min(k,l)$.  If $k \not = l$, say $k < l$, we then have 
\[t_k p_k = s_k q_k + \sum_{j = k+1}^{l} s_j q_j = t_k p_k + \sum_{j=k+1}^{l} s_j q_j\]
whence, $\sum^{l}_{j = k+1} s_j q_j = 0$, which is impossible since all $q_j$ are sharp and the coefficients $s_j$ are 
strictly positive. 
Hence, $l=k$ and the proof is complete. $\Box$ 


\section{Bits are balls}

In most other reconstructions of QM \cite{Dakic-Brukner, Masanes-Mueller, CDP}, the first step is to show that the state space of a {\em bit} --- that is, a system in which every state is the mixture of 
two sharply-distinguishable pure states --- is a ball. In our approach, this fact is an easy consequence of Lemma 1.  In our framework, we will define a bit to be 
a sharp model $A$ with uniform rank $2$, in which every state has the form $t \delta_x + (1 - t)\delta_y$, where $\{x,y\} \in \M(A)$.  Note that this implies 
$A$ is spectral. 

{\bf Lemma C1:} {\em Let $A$ be a bit with conjugate $\bar{A}$. Then $\Omega(A)$ is a ball, the extreme points of which are the states $\delta_x$, $x \in X(A)$. }

{\em Proof:} By Lemma 1, $\E(A)$ carries a self-dualizing inner product such that $\|\x\| = 1/n$ for each outcome $x \in X(A)$ and $\|u\| = 1$. Also, 
if $\{x,y\} \in \M(A)$, then $\langle \x, \y \rangle = 0$, whence, $\langle u, \x \rangle = \langle \x, \x \rangle = 1/n$. It will be convenient to adjust the normalization 
so that $\langle \x, \x \rangle  = 1$ for outcomes $x \in X(A)$, whence, $\langle u, u \rangle = n$. We can now represent states by 
vectors $a \in \E(A)_+$ with $\langle a, u \rangle = 1$. In particular, the maximally mixed state corresponds to the vector $\frac{1}{n} u$. To simplify the notation, 
let us agree for the moment to write $\rho$ for this vector. Then 
\[\langle \rho, \x \rangle = \frac{1}{n} \ \mbox{and} \  \langle \rho, \rho \rangle = \frac{1}{n^2} \langle u, u \rangle = \frac{1}{n}.\]
 Hence, 
\[\|\rho - \x\|^2 = \|\rho\|^2 - 2\langle \rho, \x \rangle + \|\x\|^2 = \frac{1}{n} - 2 \frac{1}{n}  + 1 = 1 - \frac{1}{n}.\]
If $n = 2$, we see that $\|\rho - \x\| = 1/\sqrt{2}$. Thus, $\X(A) := \{ \x | x \in X(A)\}$ lies on the sphere of radius $1/\sqrt{2}$ about the unit $\rho$. I now claim that 
any $a \in \E(A)$ with $\langle a, u \rangle = 1$ --- in effect, any state --- such that $\|u - a\| \leq 1/\sqrt{2}$,  belongs to the positive cone $\E(A)_+$.  To see this, use spectrality to decompose $a$ as 
$s \x + t \y$ where $\{x,y\} \in \M(A)$.  Consider now the two-dimensional subspace $\E_{x,y}$ spanned by $\x$ and $\y$. With respect to the inner product inherited 
from $\E$, we can regard this as a $2$-dimensional euclidean space,  in which 
$a$ is represented by the Cartesian coordinate pair $(s,t)$. Expanding $\rho$ as $\rho = \frac{1}{2}(\x + \y)$, we see that $\rho \in \E_{x,y}$ with coordinates $(1/2,1/2)$. The point 
$(t,s)$ lies, therefore, in the disk of radius $1/\sqrt{2}$ centered at $(1/2, 1/2)$ in $\E_{x,y}$. Moreover, as $\langle u, a \rangle = 1$, we see that 
$s + t = 1$, i.e., $(s,t)$ lies on the line of slope $-1$ through $(1/2,1/2)$.  This puts $(s,t)$ in the positive quadrant of this plane, i.e., $s \geq 0$ and $t \geq 0$. But 
then $a \in \E_+$, as claimed.  $\Box$ 

It follows that, for rank-two models, we do not even need to invoke homogeneity: they all correspond to spin factors. Letting $d$ denote the dimension of the state space 
(that is, $d = \dim(\E) - 1$), we see that if $d = 1$, we have the clasical bit; $d = 2$ gives the real quantum-mechanical bit, $d = 3$ gives the familiar Bloch sphere, i.e., 
the usual qubit of complex QM, while $d = 5$ corresponds to the quaternionic unit sphere, giving us the quaternionic bit.  The generalized bits with $d = 4$ and $d \geq 6$ 
are  more exotic ``post-quantum" possibilities. 

\section{Locally tomography and dagger-compactness}

\newcommand{\veta}{\vec{\eta}}
\newcommand{\valpha}{\vec{\alpha}}

A {\em dagger} on a category $\Cat$ is a contravariant functor $\dagger : \Cat \rightarrow \Cat$ that is the identity on objects, and 
satisfies $\dagger \circ \dagger = \id_{\Cat}$. That is, if $A \stackrel{f}{\longrightarrow} B$ is a morphism in $\Cat$, then $A \stackrel{f^{\dagger}}{\longleftarrow} B$, 
with $f^{\dagger \dagger} = f$ and $(f \circ g)^{\dagger} = g^{\dagger} \circ f^{\dagger}$ whenever $f \circ g$ is defined. An isomorphism 
$f : A \simeq B$ in $\Cat$ is then said to be {\em unitary} iff $f^{\dagger} = f^{-1}$. One says that $\Cat$ is {\em $\dagger$-monoidal} 
iff $\Cat$ is equipped with a symmetric monoidal structure $\otimes$ such that $(f \otimes g)^{\dagger} = f^{\dagger} \otimes g^{\dagger}$, and such that 
the canonical isomorphisms $\alpha_{A,B,C} $, $\sigma_{A,B}$, $\lambda_A$ and $\rho_A$ are all unitary. 

\newpage
A {\em dual} for an object $A$ in a symmetric monoidal category $\Cat$ is a structure $(A',\eta, \epsilon)$ where $A' \in \Cat$ and $\eta : I \rightarrow A \otimes A'$ 
and $\epsilon : A' \otimes A \rightarrow I$, such that 
\[(\id_{A} \otimes  \epsilon) \circ (\eta \otimes \id_{A}) = \id_{A} \ \ \mbox{and} \ \ (\epsilon \otimes \id_{A'}) \circ (\id_{A'} \otimes \eta) = \id_{A'}\]
up to the natural associator and unit isomorphisms.  If $\Cat$ is $\dagger$-monoidal and $\epsilon = \sigma_{A, A'} \circ \eta_{A}^{\dagger}$, then $(A', \eta, \epsilon)$ is a {\em dagger-dual}. 
A category in which every object $A$ has a specified dual $(A',\eta_A, \epsilon_A)$ is 
{\em compact closed}, and a dagger-monoidal category in which every object has a given dagger-dual is {\em dagger-compact}. See \cite{Abramsky-Coecke, Selinger} for details.

An important example of all this is the category --- I'll denote it by $\FdHilbR$ --- of finite-dimensional real Hilbert spaces and linear mappings. If $A$ and $B$ are two such spaces and $\phi : A \rightarrow B$, let 
$\phi^{\dagger}$ be the usual adjoint of $\phi$ with respect to the given inner products. Letting $A \otimes B$ be the  usual tensor product of $A$ and $B$ (in particular, with $\langle x \otimes y, u \otimes v \rangle = \langle x, u \rangle \langle y, v \rangle$ for 
$x, u \in A$ and $y, v \in B$) , $\FdHilbR$ is a dagger-monoidal category with $\R$ as the monoidal unit. 

Since any $A \in \FdHilbR$ is canonically isomorphic to its dual space, we have also a canonical isomorphism 
$A \otimes A \simeq A^{\ast} \otimes A = \L(A,A)$, and a canonical trace functional 
$\tr_{A} : A \otimes A \rightarrow \R$, uniquely defined by $\tr_{A}(x \otimes y) = \langle x, y \rangle$ 
for all $x, y \in A$. 
Taking $A' = A$, let $\eta_{A} \in A \otimes A$ be given by $\eta_{A} = \sum_{i} x_i \otimes x_i$, 
where the sum is taken over any orthonormal basis $\{x_i\}$ for $A$; then for any $a \in A \otimes A$, $\langle \eta_{A}, a \rangle = \Tr(a)$.  It is routine to show that $\tr_A  = \sigma_{A,A} \circ \eta_{A}^{\dagger}$, 
so that $\eta_A = \Tr_A$ and $\tr_A$ make $A$ its own dagger-dual.  

In any compact closed symmetric monoidal category $\Cat$, every morphism $\phi : A \rightarrow B$ yields a {\em dual} morphism $\phi' : B' \rightarrow A'$ by setting 
\[\phi' = (\id_{A'} \otimes \epsilon_{B}) \circ (\id_{A'} \otimes f \otimes \id_{B'}) \circ (\eta_{A} \otimes \id_{B'}).\]
(again, suppressing associators and left and right units). 
For $\phi : A \rightarrow B$ in $\FdHilbR$, one has, for any $v \in A$, 
\[\phi'(v) = \sum_{x \in M} \langle v, f(x) \rangle x = \sum_{x \in M} \langle f^{\dagger}(v), x \rangle x = f^{\dagger}(v),\]
i.e., $\phi' = \phi^{\dagger}$. 


Now let $\Cat$ be a monoidal probabilistic theory --- that is, a category of probabilistic models and processes, with 
a symmetric monoidal structure 
$A, B \mapsto AB$, where $AB$ is a (non-signaling) composite in our sense (see Section 6)\footnote{That is, 
$AB$ is a probabilistic model, and there is a given mapping $\pi_{AB} : X(A) \times X(B) \rightarrow \V(AB)_{+}^{\ast}$ 
with $\sum_{x \in E, y \in F} \pi(x,y) = u_{AB}$, such that for any state $\omega \in \Omega(AB)$, 
the pullback $\omega\circ \pi$ is a non-signaling joint state on $A$ and $B$.  Below, I will 
$x \otimes y$ for $\pi(x,y)$, where $x \in X(A)$ and $y \in X(B)$. Notice that $x \otimes y$ is not 
required to be an element of $X(AB)$, but rather, a positive linear functional on $\V(AB)$.  
In spite of this, I will abuse notation a bit and write $\omega(x \otimes y)$ rather than $(x \otimes y)(\omega)$ 
where $\omega \in \V(AB)$.} and with tensor unit $I = \R$.  
I will further assume that  
\begin{itemize}
\item[(a)] Every $A \in \Cat$ has a conjugate, $\bar{A} \in \Cat$, with $\bar{\bar{A}} = A$; 
\item[(b)] For all $A, B \in \Cat$ and $\phi \in \Cat(A,B)$, $\bar{\phi} \in \Cat(\bar{A}, \bar{B})$; 
\item[(c)] $\bar{\bar{A}} = A$, with $\eta_{\bar{A}}(\bar{a}, b) := \eta_{A}(a, \bar{b})$. 
\end{itemize}  

{\em Remarks:} 
(1)  The chosen conjugate $\bar{A}$ for $A \in \Cat$ required by condition (a) is equipped with a canonical isomorphism 
$\gamma_{A} : A \simeq \bar{A}$, with $\bar{x} = \gamma(x)$ for every $x \in X(A)$. As discussed in 
Section 4, this extends to an order-isomorphism $\E(A) \simeq \E(\bar{A})$, which 
we again write as $\gamma_{A}(a) = \bar{a}$ for $a \in \E(A)$.  Notice, however, that $\gamma_{A}$ 
is {\em not} assumed to be a morphism in $\Cat$. 

(2) In spite of this, condition (b) requires that $\bar{\phi} = \gamma_{B} \circ \phi \circ \gamma_{A}^{-1}$ 
{\em does} belong to  $\Cat(\bar{A},\bar{B})$  for $\phi \in \Cat(A,B)$. Notice here that $\phi \mapsto \bar{\phi}$ is functorial.  

(3)  The second part of condition (c) is redundant if every model $A$ in $\C$ is sharp (since in this case there is 
at most one correlator between $\bar{A}$ and $A$). Notice, too, that condition (c) implies that 
\[\langle \bar{x}, \bar{y} \rangle = \eta_{\bar{A}}(\bar{x}, y) = \eta_{A} (x, \bar{y}) = \langle \x, \y \rangle\]
for all $x, y \in \E(A)$.


We are now ready to prove Theorem 3.   Let $\Cat$ be a locally tomographic monoidal probabilistic theory. 
We wish to show that if every $A \in \Cat$ is sharp and spectral, then $\Cat$ has a canonical dagger, with respect to which it is dagger compact. 

Before proceeding, it will be convenient to dualize our representation of morphisms,  
so that $\phi \in \Cat(A,B)$ means that $\phi$ is a positive linear mapping 
$\E(B) \rightarrow \E(A)$.\footnote{Thus, our co-unit $\eta \in \Cat(I, A \otimes A')$  becomes a positive linear mapping $\eta_{A} : \E(A \otimes A') \rightarrow \R$, and similarly, a unit $\epsilon_{A} \in \Cat(A' \otimes A, I)$ becomes a positive linear mapping $\R \rightarrow \E(A' \otimes A)$, i.e, an element of $\E(A \otimes A')$.}
By Lemma 1, for every $A \in \Cat$, the space $\E(A)$ carries a canonical self-dualizing 
inner product $\langle \, , \, \rangle_A$, with respect to which $\E(A) \simeq \V(A)$.

{\bf Lemma D1:} {\em  For all models $A, B \in \Cat$, the inner product on $\E(AB)$ factors, in the sense that 
if $a, x \in \E(A)$ and $b, y \in \E(B)$, then $\langle a \otimes b, x \otimes y \rangle = 
\langle a, x \rangle \langle b, y \rangle$.}

{\em Proof:} This follows from the sharpness 
of $A, B$ and $AB$. For $u \in X(A)$, $v \in X(B)$, let $\delta_u, \delta_v$ and $\delta_{u \otimes v}$ denote the unique states of $A$, $B$ and $AB$ such that 
$\delta_{u}(u) = \delta_{v}(v) = \delta_{u \otimes v}(u \otimes v) = 1$.  Since $(\delta_{u} \otimes \delta_{v})(u \otimes v)$ is also $1$, we conclude that 
$\delta_{u \otimes v} = \delta_{u} \otimes \delta_{v}$. But we also have $\delta_{u}(x) = n \langle \u, \x \rangle$, $\delta_{v}(y) = m \langle \v, \y \rangle$ 
and $\delta_{u \otimes v}(x \otimes y) =  nm \langle \u \otimes \v, \x \otimes \y \rangle$, where $n, m$ and $nm$ 
are the ranks, respectively, of $A$, $B$, and $A \otimes B$. This establishes the claim. 
$\Box$ 


It follows that $\Cat$ is a monoidal subcategory of $\FdHilbR$. In effect, we are going to 
show that $\Cat$ inherits a dagger-compact structure from $\FdHilbR$, with the minor twist that 
we will take $\bar{A}$, rather than $A$, as the dual for $A \in \Cat$. We define the dagger of $\phi \in \Cat(A,B)$  to be the hermitian adjoint of $\phi : \E(A) \rightarrow \E(B)$ with respect to the canonical inner products on $\E(A)$ and $\E(B)$. At this point, it is not obvious that $\phi^{\dagger}$ belongs 
to $\Cat$.  In order to show that it does, we first need to show that $\Cat$ is compact closed. 
 To define the unit, let $e_{A} \in \E(\bar{A}) \otimes \E(\bar{A}) = \E(\bar{A} A)$ (note the use of local tomography here) 
to be the vector with $\langle e_A,  \, \cdot \, \rangle = \eta_A$, i.e, for all $a, b \in \E(A)$, 
\[\langle e_A, \bar{a} \otimes b) = \eta_{A}(a \otimes \bar{b}) = \langle a, b \rangle.\]
Since $\E(A \bar{A})$ is self-dual, $e_A \in \E(A \bar{A})_{+}$. 

{\bf Lemma D2:} {\em With $\eta_A$ and $e_{A}$ defined as above, $\bar{A}$ is a dual for $A$ for 
every $A \in \Cat$. In particular, $\Cat$ is compact closed.} 

{\em Proof:} Choose an orthonormal basis $M \subseteq \E(A)$. Local tomography and Lemma D1 tell us 
that 
$\bar{M} \otimes M = \{ \bar{a} \otimes a | a \in M\}$ is then an orthonormal basis for $\E(\bar{A}A)$. (Note here that $a, b \in M$ are not necessarily even positive, let alone in $X(A)$.)  If we expand 
$e_{A}$ with respect to this basis, we have 
\[e_{A} = \sum_{a , b \in M} \langle e_A, \bar{a} \otimes b \rangle {\bar a} \otimes b\]
Since the basis is orthonormal, we have 
\[\langle e_{A}, \bar{a} \otimes a \rangle = \langle a, a \rangle = \|a\|^2 = 1\]
and for $a \not = b$, both in $M$, 
\[\langle e_{A}, \bar{a} \otimes b \rangle = \langle a, b \rangle = 0\]
Hence, $e_{A} = \sum_{a \in M} \bar{a} \otimes a$. We now have, for any $v \in \E(A)$, 
\begin{eqnarray*}
(\eta_{A} \otimes \id_{A}) \circ (\id_{A} \otimes e_{A})(v) 
& = & 
(\eta_{A} \otimes id_{A})\left (\sum_{x \in M} v \otimes \bar{a} \otimes a \right ) \\
& = & 
\sum_{x \in M} \eta_{A}(v \otimes \bar{a}) a \\
&= & \sum_{x \in M} 
\langle v, a \rangle a = v.
\end{eqnarray*}
Similarly, for $\bar{v} \in \bar{A}$, 
\begin{eqnarray*}
(\id_{\bar{A}} \otimes \eta_{\bar{A}})  \circ (e_{\bar{A}} \otimes \id_{\bar{A}}) (\bar{v})
& = & 
(\id_{\bar{A}} \otimes \eta_{\bar{A}})\left (\sum_{a \in M} \bar{a} \otimes a \otimes \bar{v} \right ) \\ 
& = & 
\sum_{x \in M} \bar{a} \eta_{A}(a, \bar{v})  = \sum_{a \in M} \langle a, v \rangle \bar{a} \\
& = & 
\sum_{a \in M} \langle \bar{v},\bar{a} \rangle \bar{a} = \bar{v}. \ \ \Box
\end{eqnarray*}

{\bf Lemma D3:} {\em 
 If $\phi : \E(A) \rightarrow \E(B)$ belongs to $\Cat$, then so does 
$\phi^{\dagger} : \E(B) \rightarrow \E(A)$.} 

{\em Proof:} Using the compact structure on $\Cat$ defined above, 
if $\phi : A \rightarrow B$, we construct the dual of $\bar{\phi}$: 
\[ \bar{\phi}' := (\eta_B \otimes \id_A) \circ (\id_B \otimes \bar{\phi} \otimes \id_A) \circ (\id_B \otimes e_A)  
: \E(B) \rightarrow \E(A)\]
a morphism in $\Cat$. Now, if $b \in \E(B)$, applying this last mapping gives us 
\[b \mapsto (\eta_{B} \otimes \id_{A})\left ( \sum_{a \in M} b \otimes \bar{\phi}(\bar{a}) \otimes a \right ) =
 \sum_{a \in M} \eta_{B}(b, \bar{\phi}(\bar{a}))a.\]
This last expression can be rewritten:
\[\sum_{a \in M} \eta_{B}(b, \bar{\phi}(\bar{a}))a = \sum_{a \in M} \langle b, \phi(a) \rangle a 
= \sum_{a \in M} \langle \phi^{\dagger}(b), a \rangle a = \phi^{\dagger}(b).\]
Thus, the mapping $\bar{\phi}'$ above is $\phi^{\dagger}$, which, therefore, belongs to $\Cat$.  $\Box$

Thus, $\Cat$ is a dagger, as well as a monoidal, sub category of $\FdHilbR$.  Hence, 
the associator, swap, and left- and right-unit morphisms associated with an object $A \in \Cat$ are all unitary 
(since they are unitary in $\FdHilbR$), whence, $\Cat$ is dagger-monoidal. 
To complete the proof of Theorem 3, we need to check that $\eta_{A} = e_{A}^{\dagger} \circ \sigma_{A,\bar{A}} : \E(A\bar{A}) \rightarrow \R$. 
In view of our local tomography assumption, it is enough to check this on pure tensors, where a routine 
computation gives us $e_{A}^{\dagger} (\sigma_{A,\bar{A}}(a \otimes \bar{b}))$ = $\langle e_{A}^{\dagger}(\bar{b} \otimes a), 1 \rangle_{1}$  =  $\langle \bar{b} \otimes a, e_{A} \rangle_{\bar{A}A}$ = 
$\langle a, b \rangle$ = $\eta_{A}(a \otimes \bar{b})$.  $\Box$

{\em Remark:} Given that $\Cat$ is compact closed, with $\bar{A}$ the dual of $A$, 
 the functoriality of $\phi \mapsto \bar{\phi}$ makes $\Cat$ {\em strongly} compact closed, 
in the sense of \cite{Abramsky-Coecke}. This is equivalent to dagger-compactness. 

\end{document}